\documentclass{article}

\usepackage{PRIMEarxiv}

\usepackage[utf8]{inputenc} 
\usepackage[T1]{fontenc}    
\usepackage{hyperref}       
\usepackage{url}            
\usepackage{booktabs}       
\usepackage{amsfonts}       
\usepackage{nicefrac}       
\usepackage{microtype}      
\usepackage{lipsum}
\usepackage{fancyhdr}       
\usepackage{graphicx}       
\graphicspath{{media/}}     

\usepackage{geometry}
\usepackage{cite}
\usepackage{amsmath,amssymb,amsfonts}
\usepackage{algorithmic}
\usepackage{graphicx}
\usepackage{textcomp}
\usepackage{xcolor}
\usepackage{enumitem}
\usepackage{multirow}
\usepackage{hhline}
\usepackage{mathrsfs}  
\usepackage{booktabs} 
\usepackage{subfig}
\usepackage{rotating} 
\usepackage{pdflscape} 
\usepackage{makecell}
\usepackage[symbol]{footmisc}
\usepackage[numbers]{natbib}
\usepackage{hyperref}

\newcommand{\changeurlcolor}[1]{\hypersetup{urlcolor=#1}}

\DeclareMathAlphabet{\mathscrbf}{OMS}{mdugm}{b}{n}

\definecolor{green}{rgb}{0.0, 0.5, 0.0}

\newcommand\change[1]{\textcolor{black}{#1}}
\newcommand\changezs[1]{\textcolor{black}{#1}}
\newcommand{\etal}{{et al}. }
\newcommand\changefinal[1]{\textcolor{black}{#1}}
\newcommand\changeanother[1]{\textcolor{black}{#1}}
\newcommand\minorrevision[1]{\textcolor{black}{#1}}

\definecolor{green}{rgb}{0.0, 0.5, 0.0}

\newcommand\equalcontribution{\thanks{The first two authors contributed equally to this work.}}
\title{Deep Learning in Human Activity Recognition with Wearable Sensors: A Review on Advances \equalcontribution
}

\author{
  Shibo Zhang \textsuperscript{*}\\
  Northwestern University \\
  shibozhang2015@u.northwestern.edu \\
  \And
  Yaxuan Li \textsuperscript{*}\\
  McGill University \\
  yaxuanli123@gmail.com \\
   \AND
   Shen Zhang \\
   Georgia Institute of Technology \\
   shenzhang@gatech.edu \\
   \And
   Farzad Shahabi \\
   Northwestern University \\
   farzadshahabi2024@u.northwestern.edu \\
   \And
   Stephen Xia \\
   Columbia University \\
   sx2194@columbia.edu \\
   \And
   Yu Deng \\
   Northwestern University \\
   yudeng2015@u.northwestern.edu \\
   \And
   Nabil Alshurafa \\
   Northwestern University \\
   nabil@northwestern.edu \\
}

\begin{document}
\maketitle

\begin{abstract}
Mobile and wearable devices have enabled numerous applications, including activity tracking, wellness monitoring, and human--computer interaction, that measure and improve our daily lives. Many of these applications are made possible by leveraging the rich collection of low-power sensors found in many mobile and wearable devices to perform human activity recognition (HAR). Recently, deep learning has greatly pushed the boundaries of HAR on mobile and wearable devices. This paper systematically categorizes and summarizes existing work that introduces deep learning methods for wearables-based HAR and provides a comprehensive analysis of the current advancements, developing trends, and major challenges. We also present cutting-edge frontiers and future directions for deep learning-based HAR.
\end{abstract}



\section{Introduction} 
\label{sec:intro}

Since the first Linux-based smartwatch was presented in 2000 at the IEEE International Solid-State Circuits Conference (ISSCC) by Steve Mann, who was later hailed as the ``father of wearable computing”, the 21st century has witnessed a rapid growth of wearables. For example, as of January 2020, 21\% of adults in the United States, most of whom are not opposed to sharing data with medical researchers, own a smartwatch~\cite{techcrunch2020}.

\begin{figure}[h!]
\centering 
\begin{tabular}{ccc}
\includegraphics[width=0.32\linewidth]{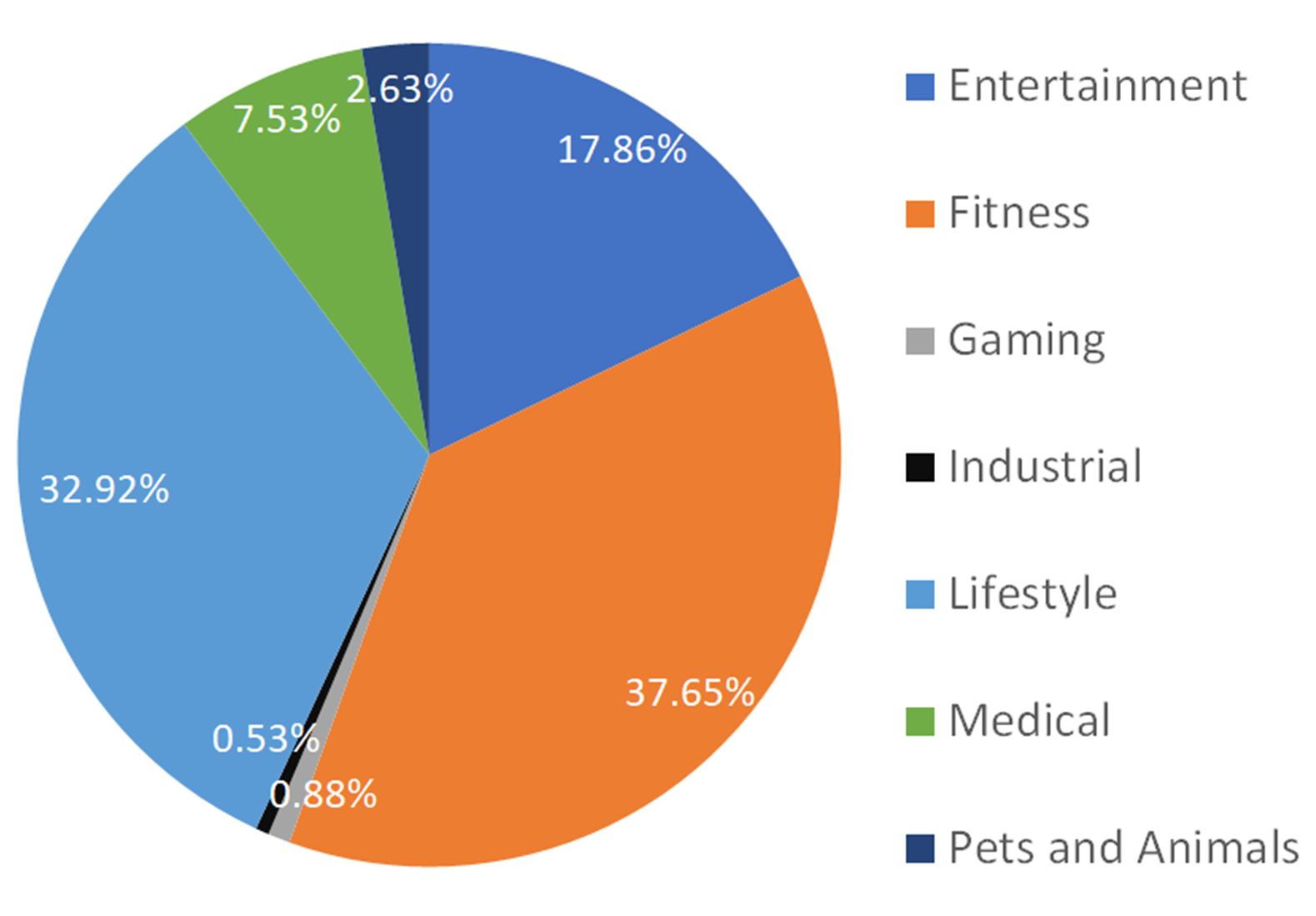}
&\includegraphics[width=0.32\linewidth]{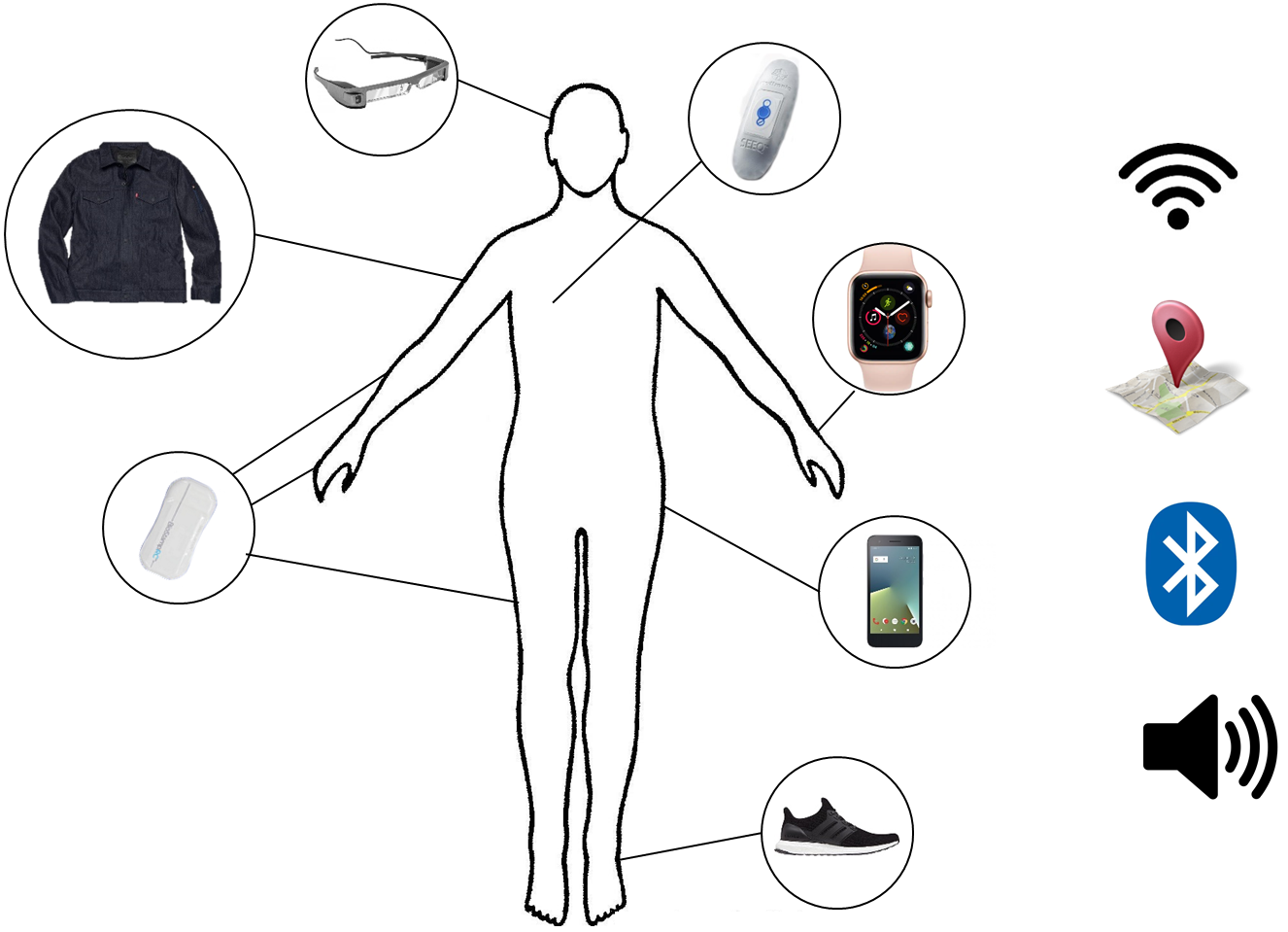}
&\includegraphics[width=0.28\linewidth]{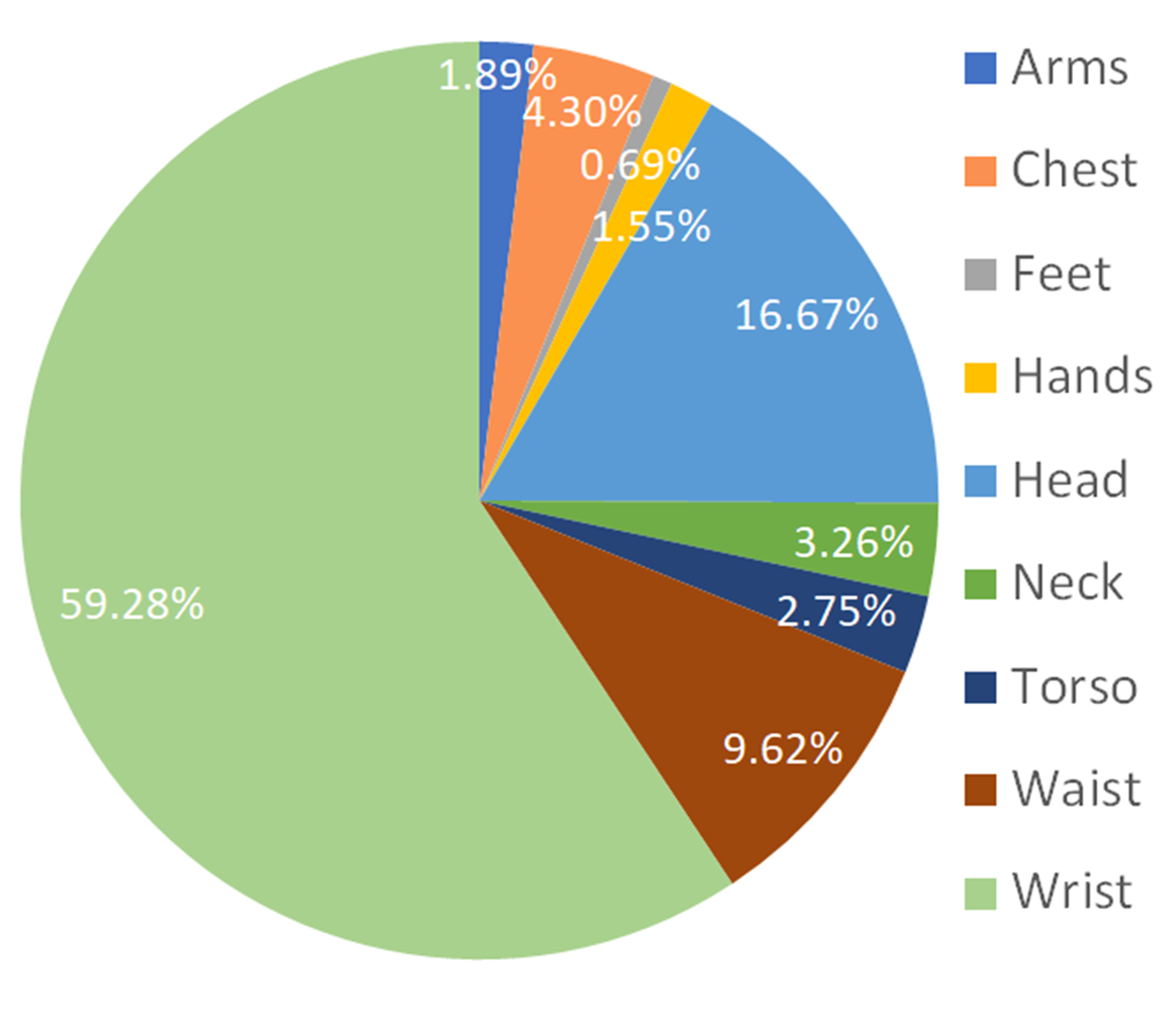}\\
({\bf a})&({\bf b})&({\bf c})\\
\end{tabular}

    \caption{Wearable devices and their application. (\textbf{a}) Distribution of wearable applications 
 \cite{wearable_database}. (\textbf{b})~Typical wearable devices. (\textbf{c}) Distribution of wearable devices placed on common body areas~\cite{wearable_database}.}\label{figg1}
\end{figure}

In addition to being fashion accessories, wearables provide unprecedented opportunities for monitoring human physiological signals and facilitating natural and seamless interaction between humans and machines. Wearables integrate low-power sensors that allow them to sense movement and other physiological signals such as heart rate, temperature, blood pressure, and electrodermal activity. The rapid proliferation of wearable technologies and advancements in sensing analytics have spurred the growth of human activity recognition (HAR). 
As a general understanding of the HAR shown in \mbox{Figure \ref{figg1}}, HAR has drastically improved the quality of service in a broad range of applications spanning healthcare, entertainment, gaming, industry, and lifestyle, among others. 
Market analysts from Meticulous Research\textregistered~\cite{wearable_forecast} forecast that the global wearable devices market will grow at a compound annual growth rate of 11.3\% from 2019, reaching \$62.82 billion by 2025, with companies like Fitbit\textregistered, Garmin\textregistered, and Huawei Technologies\textregistered ~investing more capital into the area.

In the past decade, deep learning (DL) has revolutionized traditional machine learning (ML) and brought about improved performance in many fields, including image recognition, object detection, speech recognition, and natural language processing.
DL has improved the performance and robustness of HAR, speeding its adoption and application to a wide range of wearable sensor-based applications.
%
There are two key reasons why DL is effective for many applications. First, DL methods are able to directly learn robust features from raw data for specific applications, whereas features generally need to be manually extracted or engineered in traditional ML approaches, which usually requires expert domain knowledge and a large amount of human effort. Deep neural networks can efficiently learn representative features from raw signals with little domain knowledge. Second, deep neural networks have been shown to be universal function approximators, capable of approximating almost any function given a large enough network and sufficient observations~\cite{cybenko1989approximation, scafer2006approximator, zhou2020universality}. Due to this expressive power, DL has seen \minorrevision{a substantial} growth in HAR-based applications.


Despite promising results in DL, there are still many challenges and problems to overcome, leaving room for more research opportunities. We present a review on deep learning in HAR with wearable sensors and elaborate on ongoing challenges, obstacles, and future directions in this field.

Specifically, we focus on the recognition of physical activities, including locomotion, activities of daily living (ADL), exercise, and factory work. While DL has shown a lot of promise in other applications, such as ambient scene analysis, emotion recognition, or subject identification, we focus on HAR. Throughout this work, we present brief and high-level summaries of major DL methods that have significantly impacted wearable HAR. For more details about specific algorithms or basic DL, we refer the reader to original papers, textbooks, and tutorials~\cite{Goodfellow-et-al-2016, Sutton2018}.
\change{
Our contributions are summarized as followings.}

\begin{enumerate}[label=(\roman*),labelsep=7pt]

\item \change{Firstly, we give an overview of the background of the human activity recognition research field, including the traditional and novel applications where the research community is focusing, the sensors that are utilized in these applications, as well as widely-used publicly available datasets.}
\item \change{Then, after briefly introducing the popular mainstream deep learning algorithms, we give a review of the relevant papers over the years using deep learning in human activity recognition using wearables. We categorize the papers in our scope according to the algorithm (autoencoder, CNN, RNN, etc.). In addition, we compare different DL algorithms in terms of the accuracy of the public dataset, pros and cons, deployment, and high-level model selection criteria.} 
\item \change{We provide a comprehensive \changeanother{systematic} review on the current issues, challenges, and opportunities in the HAR domain and the latest advancements towards solutions. At last, honorably and humbly, we make our best to shed light on the possible future directions with the hope to benefit students and young researchers in this field.
}
 
\end{enumerate}


\section{Methodology} 
\label{sec:method}

\subsection{Research Question} 
\label{sec:question}

\changezs{In this work, we propose several major research questions, including $\mathsf{Q1:}$ What the real-world applications of HAR, mainstream sensors, and major public datasets are in this field, $\mathsf{Q2:}$ What deep learning approaches are employed in the field of HAR and what pros and cons each of them have, and $\mathsf{Q3:}$ What challenges we are facing in this field and what opportunities and potential solutions we may have. In this work, we review the state-of-the-art work in this field and present our answers to these questions.}

This article is organized as follows:
\changezs{
We compare this work with related existing review work in this field in Section~\ref{sec:related}.} Section~\ref{sec:applications} introduces common applications for HAR. Section~\ref{sec:sensor} summarizes the types of sensors commonly used in HAR. Section~\ref{sec:dataset} summarizes major datasets that are commonly used to build HAR applications. Section~\ref{sec:deeplearning} introduces the major works in DL that contribute to HAR. Section~\ref{sec:challenges} discusses major challenges, trends, and opportunities for future work. We provide concluding remarks in Section~\ref{sec:conclusions}.

\subsection{Research Scope} 
\label{sec:scope}


%

\changeanother{In order to provide a comprehensive overview of the whole HAR field, we conducted a systematic review for human activity recognition. To ensure that our work satisfies the requirements of a high-quality systemic review, we conducted the 27-item PRISMA review process~\cite{prisma} and ensured that our work satisfied each requirement. }
We searched in Google Scholar with meta-keywords 
 ({\changefinal{We began compiling papers for this review in November 2020. As we were preparing this review, we compiled a second round of papers in November 2021 to incorporate the latest works published in 2021}}). (A) ``Human activity recognition'', ``motion recognition'', ``locomotion recognition'', ``hand gesture recognition'', ``wearable'', (B) ``deep learning'', ``autoencoder'' (alternatively ``auto-encoder''), ``deep belief network'', ``convolutional neural network'' (alternatively ``convolution neural network''), ``recurrent neural network'', ``LSTM'', ``recurrent neural network'', ``generative adversarial network'' (alternatively ``GAN''), ``reinforcement learning'', ``attention'', ``deep semi-supervised learning'', and ``graph neural network''. We used an AND rule to get combinations of the above meta-keywords (A) and (B). For each combination, we obtained top 200 search results ranked by relevance. We didn't consider any patent or citation-only search result (no content available online).



\begin{figure}[h!]\centering
  \includegraphics[width= 0.7\linewidth]{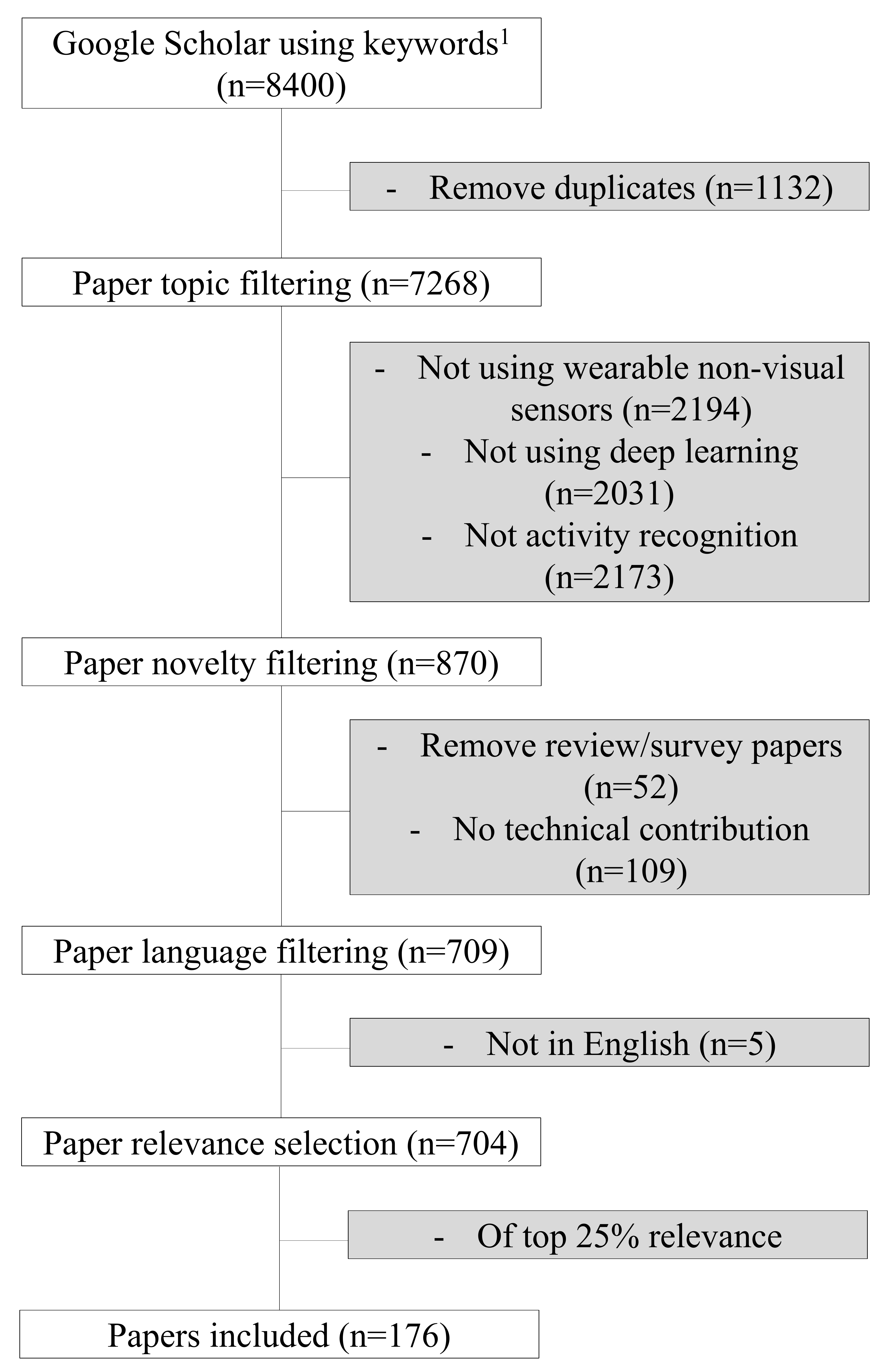}
  \caption{\change{Consort diagram outlining how we selected the final papers we included in this work.}}
  \label{fig:consort}
\end{figure}

\change{There are several exclusion criteria to build the database of the paper we reviewed.
First of all, we omitted image or video-based HAR works, such as~\cite{kiran2021fusion}, since there is a huge body of work in the computer vision community and the method is significantly different from sensor-based HAR. 
Secondly, we removed the papers using environmental sensors or systems assisted by environmental sensors such as WiFi- and RFID-based HAR. 
Thirdly, we removed the papers with minor algorithmic advancements based on prior works. We aim to present the technical progress and algorithmic achievements in HAR, so we avoid presenting works that do not stress the novelty of methods.
In the end, as the field of wearable-based HAR is becoming excessively popular and numerous papers are coming out, it is not a surprise to find that many papers share rather similar approaches, and it is almost impossible and less meaningful to cover all of them. 
\changeanother{Figure~\ref{fig:consort} shows the consort diagram that outlines step-by-step how we filtered out papers to arrive at the final 176 papers we included in this review. We obtained 8400 papers in the first step by searching keywords mentioned above on Google Scholar. Next, we removed papers that did not align with the topics in this review (i.e., works that do not utilize deep learning in wearable systems), leaving us with 870 papers. In this step, we removed 2194 papers that utilized vision, 2031 papers that did not use deep learning, and 2173 papers that did not perform human activity recognition. Then, we removed 52 review papers, 109 papers that did not propose novel systems or algorithms, and five papers that were not in English, leaving us with 704 papers. Finally, we selected the top 25\% most relevant papers to review, leaving us with 176 papers that we reviewed for this work. We used the relevancy score provided through Google Scholar to select the papers to include in this systemic review.}
Therefore, we select, categorize, and summarize representative works to present in this review paper. We adhere to the goal of our work throughout the whole paper, that is, to give an overall introduction to new researchers entering this field and present cutting-edge research challenges and opportunities.}

\changefinal{However, we admit that the review process conducted in this work has some limitations. 
%
Due to the overwhelming amount of papers in this field in recent years, it is almost impossible to include all the published papers in the field of deep learning-based wearable human activity recognition in a single review paper. The selection of the representative works to present in this paper is unavoidably subject to the risk of bias. Besides, we may miss the very first paper initiating or adopting a certain method. At last, due to the nature of human-related research and machine learning research, many possibilities could cause heterogeneity among study results, including the heterogeneity in devices, heterogeneity from the demography of participants, and even heterogeneity from the algorithm implementation details.
}

%
%
%
%



\subsection{Taxonomy of Human Activity Recognition} 
\label{sec:scope}

\begin{figure}[h!]

\centering 
  \includegraphics[width= 0.95\linewidth]{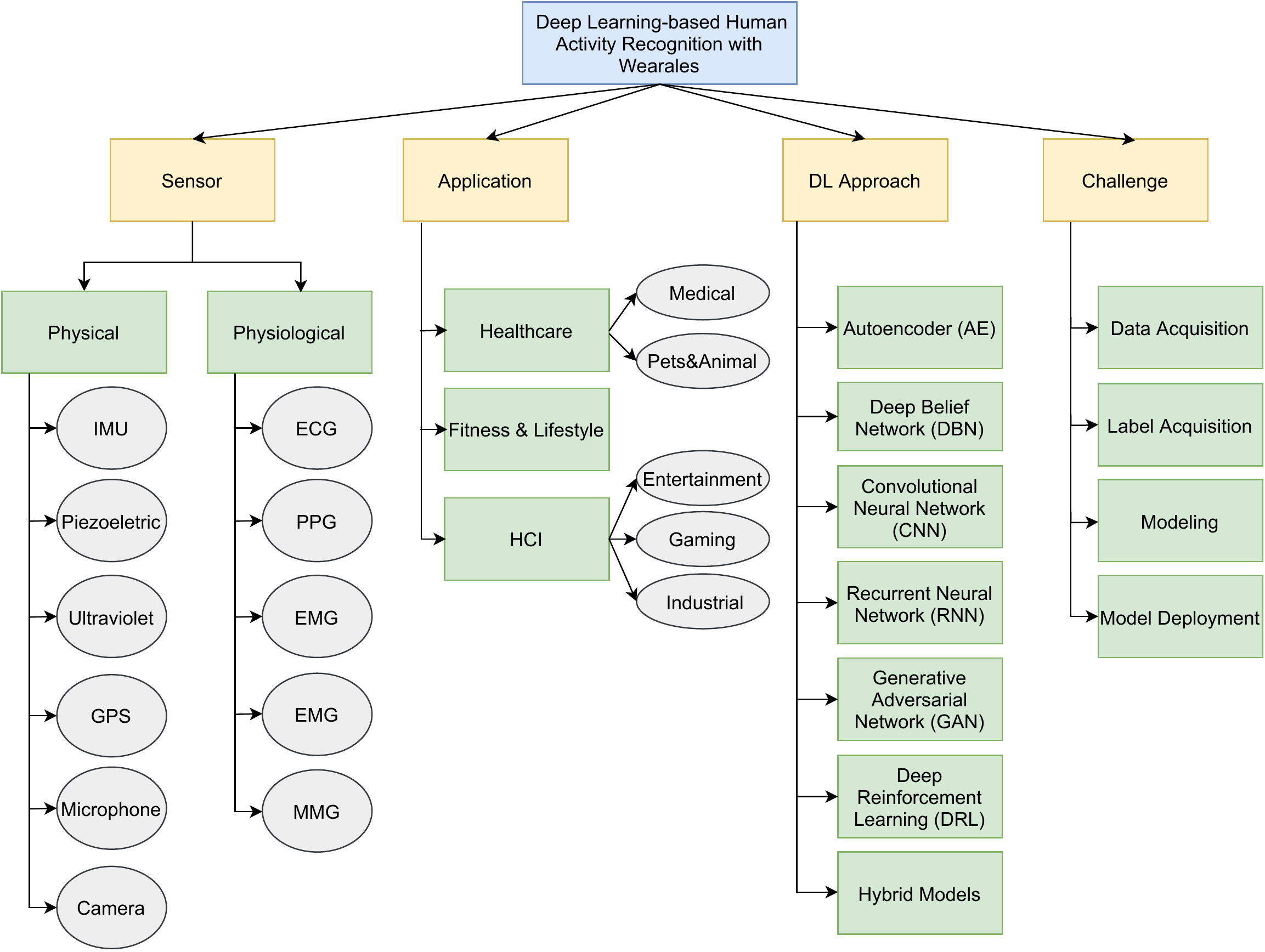}

  \caption{\change{Taxonomy of Deep Learning-based Human Activity Recognition with Wearables.}}
  \label{fig:taxonomy}
\end{figure}

\change{
In order to obtain a straightforward understanding of the hierarchies under the tree of HAR, we illustrate the taxonomy of HAR as shown in Figure~\ref{fig:taxonomy}. We categorized existing HAR works into four dimensions: Sensor, application, DL approach, and challenge. There are basically two kinds of sensors: Physical sensors and physiological sensors. Physical sensors include Inertial Measurement Unit (IMU), piezoelectric sensor, GPS, wearable camera, etc. Some exemplary physiological sensors are electromyography (EMG) and photoplethysmography (PPG), just to name a few. 
In terms of the applications of HAR systems, we categorized them into healthcare, fitness\&
lifestyle, and Human Computer Interaction (HCI). Regarding the DL algorithm, we introduce six approaches, including autoencoder (AE), Deep Belief Network (DBN), Convolutional Neural Network, Recurrent Neural Network (including Long Short-Term Memory (LSTM) and Gated Recurrent Units (GRUs)), Generative Adversarial Network (GAN), and Deep Reinforcement Learning (DRL). In the end, we discuss the challenges our research community is facing and the state-of-the-art works are coping with, also shown in Figure~\ref{fig:taxonomy}.
}
%

\section{Related Work} 
\label{sec:related}

\change{
There are some existing review papers in the literature for deep learning approaches for sensor-based human activity recognition~\cite{wang2019deep,nweke2018deep,chen2021deep,ramanujam2021human}.}
\change{
Nweke \etal accentuated the advancements in deep learning models by proposing a taxonomy of generative, discriminative, and hybrid methods along with further explorations for the advantages and limitations up to year 2018~\cite{nweke2018deep}. 
Similarly, Wang \etal conducted a thorough analysis on different sensor-based modalities, deep learning models, and their respective applications up to the year 2017~\cite{wang2019deep}. 
However, in recent years, due to huge advancements in the availability and computational power of computing resources and cutting-edge deep learning techniques, the applied deep learning area has been revolutionized and reached all-time-high performance in the field of sensor-based human activity recognition. Therefore, we aim to present the most recent advances and most exciting achievements in the community in a timely manner to our readers.}

\change{In another work, Chen \etal provided the community a comprehensive review which has done an in-depth analysis of the challenges/opportunities for deep learning in sensor-based HAR and proposed a new taxonomy for the challenges ahead of the activity recognition systems~\cite{chen2021deep}. 
In contrast, we view our work as more of a gentle introduction of this field to students and novices in the way that our literature review provides the community with a detailed analysis on most recent state-of-the-art deep learning architectures (i.e., CNN, RNN, GAN, Deep Reinforcement Learning, and hybrid models) and their respective pros and cons on HAR benchmark datasets. At the same time, we distill our knowledge and experience from our past works in this field and present the challenges and opportunities from a different viewpoint.
Another recent work was presented by Ramanujam \etal, in which they categorized the deep learning architectures in CNN, LSTM, and hybrid methods and conducted an in-depth analysis on the benchmark datasets~\cite{ramanujam2021human}. 
Compared with their work, our paper pays more attention to the most recent cutting-edge deep learning methods applied on HAR on-body sensory data, such as GAN and DRL. We also provide both new learners and experienced researchers with a profound resource in terms of model comparison, model selection and model deployment.
%
%
%
In a nutshell, our review has thoroughly analysed most up-to-date deep learning architectures applied on various wearable sensors, elaborated on their respective applications, and compared performances on public datasets. What's more, we attempt to cover the most recent advances in resolving the challenges and difficulties and shed light on possible research opportunities. 
}



\section{Human Activity Recognition Overview}

\subsection{Applications} \label{sec:applications}
%

In this section, we illustrate the major areas and applications of wearable devices in HAR. Figure~\ref{figg1}a, taken from the wearable technology database~\cite{wearable_database}, breaks down the distribution of application types of 582 commercial wearables registered since 2015~\cite{wearable_database}. The database suggests that wearables are increasing in popularity and will impact people's lives in several ways, particularly in applications ranging from fitness and lifestyle to medical and human-computer interaction.

\subsubsection{Wearables in Fitness and Lifestyle}
Physical activity involves activities such as sitting, walking, laying down, going up or downstairs, jogging, and running~\cite{Physical_activity_recognition}. 
Regular physical activity is increasingly being linked to a reduction in risk for many chronic diseases, such as obesity, diabetes, and cardiovascular disease, and has been shown to improve mental health~\cite{mobility_chronicdisease}. 
The data recorded by wearable devices during these activities include plenty of information, such as duration and intensity of activity, which further reveals an individual's daily habits and health conditions~\cite{physical_activity_health}. 
For example, dedicated products such as Fitbit~\cite{fitbit} can estimate and record energy expenditure on smart devices, which can further serve as an important step in tracking personal activity and preventing chronic diseases~\cite{Zhu2015Using}. Moreover, there has been evidence of the association between modes of transport (motor vehicle, walking, cycling, and public transport) and obesity-related outcomes~\cite{transport_medical}.
Being aware of daily locomotion and transportation patterns can provide physicians with the necessary information to better understand patients' conditions and also encourage users to engage in more exercise to promote behavior change~\cite{Fitness_Behavior_Change}. Therefore, the use of wearables in fitness and lifestyle has the potential to significantly advance one of the most prolific aspects of HAR \mbox{applications~\cite{cnnhar2014, cnn2015smc, cnn2015mm, cnn2016expert, cnn2017bigcomp, benchmarkingSHL, SHLIndRNN, TMD2018, Attention_CNN_weaklylabel}.}

Energy (or calorie) expenditure (EE) estimation has grown to be an important reason why people care to track their personal activity. Self-reflection and self-regulation of one's own behavior and the habit has been important factor in designing interventions that prevent chronic diseases such as obesity, diabetes, and cardiovascular diseases. 

\subsubsection{Wearables in Healthcare and Rehabilitation}

HAR has greatly impacted the ability to diagnose and capture pertinent information in healthcare and rehabilitation domains. 
By tracking, storing, and sharing patient data with medical institutions, wearables have become instrumental for physicians in patient health assessment and monitoring. 
Specifically, several works have introduced systems and methods for monitoring and assessing  Parkinson disease (PD) symptoms~\cite{parkinson2015, parkinson2016, parkinson2017, parkinson2018, parkinson2018_2, parkinson_augmentation}. 
\change{Pulmonary disease, such as Chronic Obstructive Pulmonary Disease (COPD), asthma, and COVID-19, is one of leading causes of morbidity and mortality. Some recent works use wearables to detect cough activity, a major symptom of pulmonary diseases~\cite{Listen2Cough, mcc,CoughBuddy2021Nemati,zhang2021coughtrigger}}.
Other works have introduced methods for monitoring stroke in infants using wearable accelerometers~\cite{infant_stroke} and methods for assessing depressive symptoms utilizing wrist-worn sensors~\cite{depression}.
In addition,  detecting muscular activities and hand motions using electromyography (EMG) sensors has been widely applied to enable improved prostheses control for people with missing or damaged limbs~\cite{EMG_review, EMG_armband, EMG_prosthese, EMG_prosthese2}.




\subsubsection{Wearables in Human Computer Interaction (HCI)}

Modern wearable technology in HCI has provided us with flexible and convenient methods to control and communicate with electronics, computers, and robots. 
For example, a wrist-worn wearable outfitted with an inertial measurement unit (IMU) can easily detect the wrist shaking~\cite{seg_cnn, dynamichandgesturerecognition, fingerping} to control smart devices to skip a song by shaking the hand, instead of bringing up the screen, locating, and pushing a button. 
Furthermore, wearable devices have played an essential role in many HCI applications in entertainment systems and immersive technology.
One example field is augmented reality (AR) and virtual reality (VR), which has changed the way we interact and view the world. Thanks to accurate activity, gesture, and motion detection from wearables, these applications could induce feelings of cold or hot weather by providing an immersive experience by varying the virtual environment and could enable more realistic interaction between the human and virtual objects~\cite{EMG_review, EMG_armband}.

\subsection{Wearable Sensors}
\label{sec:sensor}
Wearable sensors are the foundation of HAR systems. As shown in Figure~\ref{figg1}b, there are a large number of off-the-shelf smart devices or prototypes under development today, including smartphones, smartwatches, smart glasses, smart rings~\cite{ring}, smart gloves~\cite{nature_gloves}, smart armbands~\cite{armband}, smart necklaces~\cite{necklace2015, necksense, NeckFace2021Chen}, smart shoes~\cite{shoes}, and E-tattoos~\cite{e-tattoo}. These wearable devices cover the human body from head to toe with a general distribution of devices shown in Figure~\ref{figg1}c, as reported by~\cite{wearable_database}.
The advance of micro-electro-mechanical system (MEMS) technology (microscopic devices, comprising a central unit such as a microprocessor and multiple components that interact with the surroundings such as microsensors) has allowed wearables to be miniaturized and lightweight to reduce the burden on adherence to the use of wearables and Internet of Things (IoT) technologies. In this section, we introduce and discuss some of the most prevalent MEMS sensors commonly used in wearables for HAR.
\change{The summary of wearable sensors is represented as a part of Figure \ref{fig:taxonomy}.}



\subsubsection{Inertial Measurement Unit (IMU)}
Inertial measurement unit (IMU) is an integrated sensor package comprising of accelerometer, gyroscope, and sometimes magnetometer. Specifically, an accelerometer detects linear motion and gravitational forces by measuring the acceleration in 3 axes (x, y, and z), while a gyroscope measures rotation rate (roll, yaw, and pitch). The magnetometer is used to detect and measure the earth's magnetic fields. Since a magnetometer is often used to obtain the posture and orientation in accordance with the geomagnetic field, which is typically outside the scope of HAR, the magnetometer is not always included in data analysis for HAR.  
By contrast, accelerometers and gyroscopes are commonly used in many HAR applications. We refer to an IMU package comprising a 3-axis accelerometer and a 3-axis gyroscope as a 6-axis IMU. This component is often referred to as a 9-axis IMU if a 3-axis magnetometer is also integrated. Owing to mass manufacturing and the widespread use of smartphones and wearable devices in our daily lives, IMU data are becoming more ubiquitous and more readily available to collect. 
In many HAR applications, researchers carefully choose the sampling rate of the IMU sensors depending on the activity of interest, often choosing to sample between 10 and several hundred Hz. 
In~\cite{SampleRate}, Chung \etal  tested a range of sampling rates and gave the best one in his application. 
Besides, it's been shown that higher sampling rates allow the system to capture signals with higher precision and frequencies, leading to more accurate models at the cost of higher energy and resource consumption. 
For example, the projects presented in~\cite{ViBand, Laput2} utilize sampling rates above the typical rate. These works sample at 4 kHz to sense the vibrations generated from the interaction between a hand and a physical object. 

%

\subsubsection{Electrocardiography (ECG) and Photoplethysmography (PPG)}

Electrocardiography (ECG) and photoplethysmography (PPG) are the most commonly used sensing modalities for heart rate monitoring. ECG, also called EKG, detects the heart's electrical activity through electrodes attached to the body. The standard 12-lead ECG attaches ten non-intrusive electrodes to form 12 leads on the limbs and chest. ECG is primarily employed to detect and diagnose cardiovascular disease and abnormal cardiac rhythms. PPG relies on using a low-intensity infrared (IR) light sensor to measure blood flow caused by the expansion and contraction of heart chambers and blood vessels. Changes in blood flow are detected by the PPG sensor as changes in the intensity of light; filters are then applied to the signal to obtain an estimate of heart rate. 
Since ECG directly measures the electrical signals that control heart activity, it typically provides more accurate measurements for heart rate and often serves as a baseline for evaluating PPG sensors.

%

\subsubsection{Electromyography (EMG)}
Electromyography (EMG) measures the electrical activity produced by muscle movement and contractions. EMG was first introduced in clinical tests to assess and diagnose the functionality of muscles and motor neurons. There are two types of EMG sensors: Surface EMG (sEMG) and intramuscular EMG (iEMG). sEMG uses an array of electrodes placed on the skin to measure the electrical signals generated by muscles through the surface of the skin~\cite{wiki:emg}. 
There are a number of wearable applications that detect and assess daily activities using sEMG~\cite{EMG_armband, EMG_handmotion}. 
In~\cite{cnn12}, researchers developed a neural network that distinguishes ten different hand motions using sEMG to advance the effectiveness of prosthetic hands.
iEMG places electrodes directly into the muscle beneath the skin. Because of its invasive nature, non-invasive wearable HAR systems do not typically include iEMG.

\subsubsection{Mechanomyography (MMG)}
Mechanomyography (MMG) uses a microphone or accelerometer to measure low-frequency muscle contractions and vibrations, as opposed to EMG, which uses electrodes. For example, 4-channel MMG signals from the thigh can be used to detect knee motion patterns~\cite{cnn11}. Detecting these knee motions is helpful for the development of power-assisted wearables for powered lower limb prostheses. The authors create a convolutional neural network and support vector machine (CNN-SVM) architecture comprising a seven-layer CNN to learn dominant features for specific knee movements. The authors then replace the fully connected layers with an SVM classifier trained with the extracted feature vectors to improve knee motion pattern recognition.
Moreover, Meagher \etal~\cite{meagher2020new} proposed developing an MMG device as a wearable sensor to detect mechanical muscle activity for rehabilitation after stroke.

%

Other wearable sensors used in HAR include (but are not limited to) \change{piezoelectric sensor~\cite{khalifa2017harke, cha2018flexible} for converting changes in pressure, acceleration, temperature, strain, or force to electrical charge, barometric pressure sensor~\cite{masse2015improving} for atmospheric pressure, temperature measurement~\cite{barna2019study},} electroencephalography (EEG) for measuring brain activity~\cite{salehzadeh2020human}, respiration sensors for breathing monitoring~\cite{ramos2016using}, ultraviolet (UV) sensors~\cite{filippoupolitis2016activity} for sun exposure assessment, GPS for location sensing, \change{microphones for audio recording~\cite{jin2021sonicasl, ntalampiras2018transfer, CoughBuddy2021Nemati}}, and wearable cameras for image or video recording~\cite{NeckFace2021Chen}. 
%
It is also important to note that the wearable camera market has drastically grown with cameras such as GoPro becoming mainstream~\cite{GoPro1, GoPro2, alharbi2018can, alharbi2019mask} over the last few years. However, due to privacy concerns posed by participants related to video recording, utilizing wearable cameras for longitudinal activity recognition is not as prevalent as other sensors. Additionally, HAR with image/video processing has been extensively studied in the computer vision community~\cite{voulodimos2018deep, xu2020computer}, and the methodologies commonly used differ significantly from techniques used for IMUs, EEG, PPG, etc. For these reasons, despite their significance in applications of deep learning methods, this work does not cover image and video sensing for HAR.






\subsection{Major Datasets} \label{sec:dataset}

We list the major datasets employed to train and evaluate various ML and DL techniques in Table~\ref{tab:datasets}, ranked based on the number of citations they received per year according to Google Scholar. 
As described in the earlier sections, most datasets are collected via IMU, GPS, or ECG. While most datasets are used to recognize physical activity or daily activities~\cite{Opportunity, USC-HAD, UCI-HAR, DSA, TROIKA, SHO, WISDM, WHARF, HASC, SARD, Stisen, mHealth1, mHealth2, UTD-MHAD1/2, ExtraSensory, SHAR, Actitracker, PAMAP2, Ubicomp_08}, there are also a few datasets dedicated to hand gestures~\cite{ActRecTut, FIC}, breathing patterns~\cite{BreathPrint}, and car assembly line activities~\cite{Skoda}, as well as those that monitor gait for patients with PD~\cite{Daphnet}.

\begin{table}[h!]
\centering
\small
    \caption{Major Public Datasets for Wearable-based HAR.}
    \renewcommand\tabcolsep{5.9pt}

\centering 
\begin{tabular}{lcccccc}
\toprule
\textbf{Dataset} & \textbf{Application} & \textbf{Sensor} &\textbf{ \# Classes} & \textbf{Spl. Rate} & \textbf{Citations/yr }\\
\midrule
WISDM~\cite{WISDM} & Locomotion & 3D Acc. & 6 & 20 Hz &  217 \\
ActRecTut~\cite{ActRecTut} & Hand gestures & 9D IMU & 12 & 32 Hz & 153 \\
UCR(UEA)-TSC~\cite{UCRArchive, UEA} & 9 datasets (e.g., uWave~\cite{uWave}) & Vary & Vary & Vary & 107 \\ 
UCI-HAR~\cite{UCI-HAR} & Locomotion & Smartphone 9D IMU  & 6 & 50 Hz & 78 \\
Ubicomp 08 ~\cite{Ubicomp_08} & Home activities &  Proximity sensors & 8 & N/A & 69 \\
SHO~\cite{SHO} & Locomotion & Smartphone 9D IMU & 7 & 50 Hz & 52 \\
UTD-MHAD1/2~\cite{UTD-MHAD1/2} & Locomotion \& activities & 3D Acc. \& 3D Gyro. & 27 & 50 Hz & 39 \\
HHAR~\cite{Stisen} & Locomotion & 3D Acc. & 6 & 50--200 Hz & 37 \\
Daily \& Sports Activities~\cite{DSA} & Locomotion & 9D IMU & 19 & 25 Hz & 37  \\
MHEALTH~\cite{mHealth1, mHealth2} & Locomotion \& gesture & 9D IMU \& ECG & 12 & 50 Hz & 33 \\
Opportunity~\cite{Opportunity} & Locomotion \& gesture  & 9D IMU & 16 & 50 Hz & 32 \\
PAMAP2~\cite{PAMAP2} & Locomotion \& activities & 9D IMU \& HR monitor & 18 & 100 Hz & 32 \\
Daphnet~\cite{Daphnet} & Freezing of gait & 3D Acc. & 2 & 64 Hz & 30 \\
SHL~\cite{SHL} & Locomotion \& transportation & 9D IMU & 8 & 100 Hz & 23 \\
SARD~\cite{SARD} & Locomotion & 9D IMU \& GPS & 6 & 50 Hz & 22 \\
Skoda Checkpoint~\cite{Skoda} & Assembly-line activities & 3D Acc. & 11 & 98 Hz & 21 \\
UniMiB SHAR~\cite{SHAR} & Locomotion \& gesture & 9D IMU & 12 & N/A & 20 \\
USC-HAD~\cite{USC-HAD} & Locomotion & 3D ACC. \& 3D Gyro. &  12 & 100 Hz & 20 \\
ExtraSensory~\cite{ExtraSensory} & Locomotion \& activities & 9D IMU \& GPS & 10 & 25--40 Hz & 13 \\
HASC~\cite{HASC} & Locomotion & Smartphone 9D IMU & 6 & 100 Hz & 11 \\
Actitracker~\cite{Actitracker} & Locomotion & 9D IMU \& GPS & 5 & N/A & 6 \\
FIC~\cite{FIC} & Feeding gestures  & 3D Acc. & 6 & 20 Hz & 5 \\
WHARF~\cite{WHARF} & Locomotion & Smartphone 9D IMU & 16 & 50 Hz & 4 \\
\bottomrule
\end{tabular}

\label{tab:datasets}
\end{table}

%

\begin{figure}[h!]
\centering 
  \includegraphics[width= \linewidth]{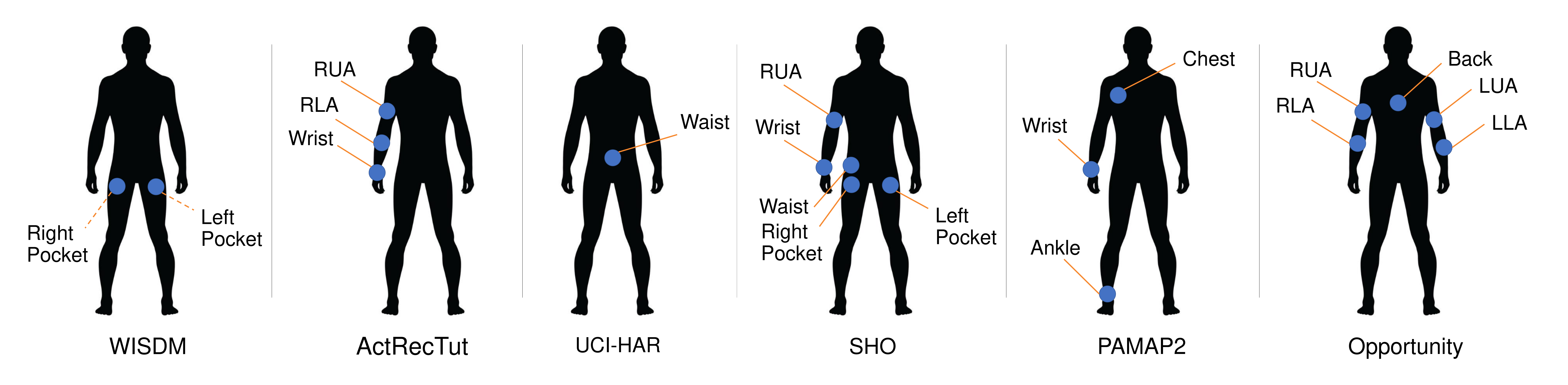}
  \caption{Placement 
 of inertial sensors in different datasets: WISDOM; ActRecTut; UCI-HAR; SHO; PAMAP2; and Opportunity.}
  \label{fig:location}
\end{figure}

Most of the datasets listed above are publicly available. The University of California Riverside-Time Series Classification (UCR-TSC) archive is a collection of datasets collected from various sensing modalities~\cite{UCR}. The UCR-TSC archive was first released and included 16 datasets, growing to 85 datasets by 2015 and 128 by October 2018. Recently, researchers from the University of East Anglia have collaborated with UCR to generate a new collection of datasets, which includes nine categories of HAR: {{BasicMotions}
}, {Cricket}, {Epilepsy}, {ERing}, {Handwriting}, {Libras}, {NATOPS}, {RacketSports}, and {UWaveGestureLibrary}~\cite{UEA}.
One of the most commonly used datasets is the OPPORTUNITY dataset~\cite{Opportunity}. This dataset contains data collected from 12 subjects using 15 wireless and wired networked sensor systems, with 72 sensors and ten modalities attached to the body or the environment. 
Existing HAR papers mainly focus on data from on-body sensors, including 7 IMUs and 12 additional 3D accelerometers for classifying 18 kinds of activities. Researchers have proposed various algorithms to extract features from sensor signals and to perform activity classification using machine-learned models like K Nearest Neighbor (KNN) and SVM~\cite{knn_har1, knn_har2, SVM_har1, cnn5, cnnhar2014, cnn17, cnn57, cnn61, cnn84, cnn115}.
Another widely-used dataset is PAMAP2~\cite{PAMAP2}, which is collected from 9 subjects performing 18 different activities, ranging from jumping to house cleaning, with 3 IMUs (100-Hz sampling rate) and a heart rate monitor (9 Hz) attached to each subject. Other datasets such as Skoda~\cite{Skoda} and WISDM~\cite{WISDM} are also commonly used to train and evaluate HAR algorithms. In Figure~\ref{fig:location}, we present the placement of inertial sensors in 9 common datasets.

\vspace{-6pt}

\section{Deep Learning Approaches}
\label{sec:deeplearning}

In recent years, DL approaches have outperformed traditional ML approaches in a wide range of HAR tasks. There are three key factors behind deep learning's success: Increasingly available data, hardware acceleration, and algorithmic advancements. The growth of datasets publicly shared through the web has allowed developers and researchers to quickly develop robust and complex models. The development of GPUs and FPGAs have drastically shortened the training time of complex and large models. Finally, improvements in optimization and training techniques have also improved training speed. In this section, we will describe and summarize HAR works from six types of deep learning approaches. 
\change{
We also present an overview of deep learning approaches in Figure \ref{fig:taxonomy}.}

%


\subsection{Autoencoder}

\begin{figure}[h!]
\centering
  \includegraphics[width=0.7\linewidth]{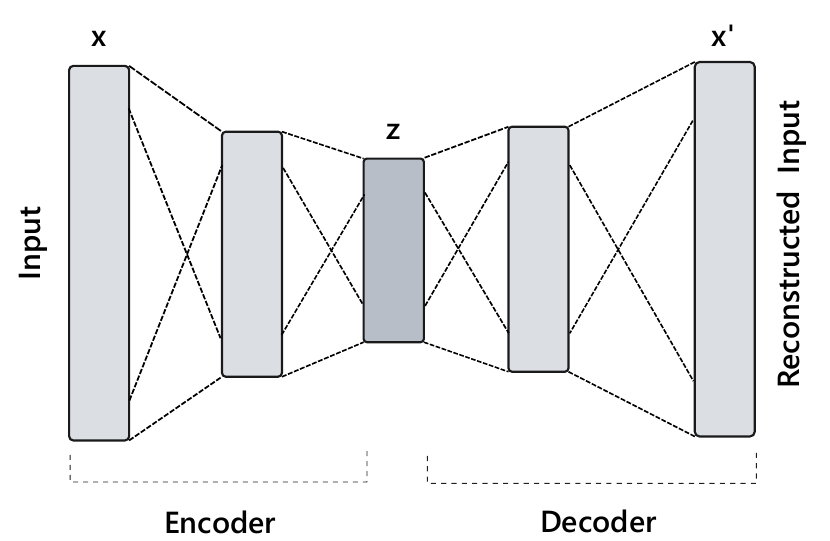}
  \caption{Illustration of an autoencoder network~\cite{autoencoder_graph}.}
  \label{fig:fig_ae}
\end{figure}

The autoencoder, originally called ``autoassociative learning module'', was first proposed in the 1980s as an unsupervised pre-training method for artificial neural networks (ANN)~\cite{AE0}.
Autoencoders have been widely adopted as an unsupervised method for learning features. As such, the outputs of autoencoders are often used as inputs to other networks and algorithms to improve performance~\cite{AE420, AE437, AE440, AE460, AE463}. An autoencoder is generally composed of an encoder module and a decoder module. The encoding module encodes the input signals into a latent space, while the decoder module transforms signals from the latent space back into the original domain. As shown in Figure \ref{fig:fig_ae}, the encoder and decoder module is usually several dense layers (i.e., fully connected layers) of the form 

$$
\begin{array}{l}
f_{\theta}(\mathbf{x}): \mathbf{z}=\sigma\left(W_{e} \mathbf{x} + b_{e}\right) \\
g_{\theta^{\prime}}(\mathbf{z}): \mathbf{x}^{\prime}=\sigma\left(W_{d} \mathbf{z} + b_{d}\right)
\end{array}
$$
where $\theta=\left\{W_{e}, b_{e}\right\}$, $\theta^{\prime}=\left\{W_{d}, b_{d}\right\}$ are the learnable parameters of the encoder and decoder. $\sigma$ is the non-linear activation function, such as Sigmoid, tanh, or rectified linear unit (ReLU). $W_{e}$ and $W_{d}$ refer to the weights of the layer, while $b_e$ and $b_d$ are the bias vectors. By minimizing a loss function applied on $\mathbf{x}$ and $\mathbf{x}^{\prime}$, autoencoders aim at generating the final output by imitating the input. Autoencoders are efficient tools for finding optimal codes, $\mathbf{z}$, and performing dimensionality reduction. An autoencoder's strength in dimensionality reduction has been applied to HAR in wearables~\cite{AE1, AE2, AE3, parkinson2018, AE437, AE6, AE7, AE8, AE423} and functions as a powerful tool for denoising and information retrieval.
\vspace{-9pt}

As such, autoencoders are most commonly used for feature extraction and dimensionality reduction~\cite{AE1, AE2, AE419, AE420, AE422, AE426, AE427, AE432, AE436, AE440, AE445, AE452, AE460, AE463, AE464}. Autoencoders are generally used individually or in a stacked architecture with multiple autoencoders.
{Mean squared error} or {mean squared error plus KL divergence} loss functions are typically used to train autoencoders. Li \etal presents an autoencoder architecture where a sparse autoencoder and a denoising autoencoder are used to explore useful feature representations from accelerometer and gyroscope sensor data, and then they perform classification using support vector machines~\cite{AE1}. Experiments are performed on a public HAR dataset~\cite{UCI-HAR} from the UCI repository, and the classification accuracy is compared with that of Fast Fourier Transform (FFT) in the frequency domain and Principal Component Analysis (PCA). The result reveals that the stacked autoencoder has the highest accuracy of 92.16\% and provides a 7\% advantage over traditional methods with hand-crafted features. Jun and Choi~\cite{AE481} studied the classification of newborn and infant activities into four classes: Sleeping, moving in agony, moving in normal condition, and movement by an external force. Using the data from an accelerometer attached to the body and a three-layer autoencoder combined with k-means clustering, they achieve 96\% weighted accuracy in an unsupervised way. 
Additionally, autoencoders have been explored for feature extraction in domain transfer learning~\cite{AE439}, detecting unseen data~\cite{AE434}, and recognizing null classes~\cite{AE435}. \change{For example, \citet{prabono2021atypical} propose a two-phase autoencoder-based approach of domain adaptation for human activity recognition. In addition, \citet{garcia2021ensemble} proposed an effective multi-class algorithm that consists of an ensemble of autoencoders where each autoencoder is associated with a separate class. This modular structure of classifiers makes models more flexible when adding new classes, which only calls for adding new autoencoders instead of re-training the model.}

Furthermore, autoencoders are commonly used to sanitize and denoise raw sensor data~\cite{AE3, AE8, AE467}, a known problem with wearable signals that impacts our ability to learn patterns in the data.
Mohammed and Tashev in~\cite{AE3} investigated the use of sensors integrated into common pieces of clothing for HAR.
However, they found that sensors attached to loose clothing are prone to contain large amounts of motion artifacts, leading to low mean signal-to-noise ratios (SNR). 
To remove motion artifacts, the authors propose a deconvolutional sequence-to-sequence autoencoder (DSTSAE). The weights for this network are trained with a weighted form of a standard VAE loss function. Experiments show that the DSTSAE outperforms traditional Kalman Filters and improves the SNR from $-$12 dB to  +18.2 dB, with the F1-score of recognizing gestures improved by 14.4\% and locomotion activities by 55.3\%. 
Gao \etal explores the use of stacking autoencoders to denoise raw sensor data to improve HAR using the UCI dataset~\cite{UCI-HAR}~\cite{AE8}. Then, LightGBM (LBG) is used to classify activities using the denoised signals.

Autoencoders are also commonly used to detect abnormal muscle movements, such as Parkinson's Disease and Autism Spectrum Disorder (ASD).
Rad \etal in~\cite{parkinson2018} utilizes an autoencoder to denoise and extract optimized features of different movements and use a one-class SVM to detect movement anomalies. To reduce the overfitting of the autoencoder, the authors inject artificial noise to simulate different types of perturbations into the training data. Sigcha \etal in~\cite{AE478} uses a denoising autoencoder to detect freezing of gait (FOG) in Parkinson's disease patients. The autoencoder is only trained using data labelled as a normal movement. During the testing phase, samples with significant statistical differences from training data are classified as abnormal FOG events.

As autoencoders map data into a nonlinear and low-dimensional latent space, they are well-suited for applications requiring privacy preservation. Malekzadeh \etal developed a novel replacement autoencoder that removes prominent features of sensitive activities, such as drinking, smoking, or using the restroom~\cite{AE437}. Specifically, the replacement autoencoder is trained to produce a non-sensitive output from a sensitive input via stochastic replacement while keeping characteristics of other less sensitive activities unchanged. Extensive experiments are performed on Opportunity~\cite{Opportunity}, Skoda~\cite{Skoda}, and \mbox{Hand-Gesture~\cite{ActRecTut}} datasets. The result shows that the proposed replacement autoencoder can retain the recognition accuracy of non-sensitive tasks using state-of-the-art techniques while simultaneously reducing detection capability for sensitive tasks. 

Mohammad \etal introduces a framework called Guardian-Estimator-Neutralizer (GEN) that attempts to recognize activities while preserving gender privacy~\cite{AE6}. 
The rationale behind GEN is to transform the data into a set of features containing only non-sensitive features. 
The Guardian, which is constructed by a deep denoising autoencoder, transforms the data into representation in an inference-specific space. 
The Estimator comprises a multitask convolutional neural network that guides the Guardian by estimating sensitive and non-sensitive information in the transformed data. Due to privacy concerns, it attempts to recognize an activity without disclosing a participant's gender. 
The Neutralizer is an optimizer that helps the Guardian converge to a near-optimal transformation function. 
Both the publicly available MobiAct~\cite{MobiAct} and a new dataset, MotionSense, are used to evaluate the proposed framework's efficacy. 
Experimental results demonstrate that the proposed framework can maintain the usefulness of the transformed data for activity recognition while reducing the gender classification accuracy to 50\% (random guessing) from more than 90\% when using raw sensor data. 
Similarly, the same authors have proposed another anonymizing autoencoder in~\cite{AE7} for classifying different activities while reducing user identification accuracy. Unlike most works, where the output to the encoder is used as features for classification, this work utilizes both the encoder and decoder outputs. Experiments performed on a self-collected dataset from the accelerometer and gyroscope showcased excellent activity recognition performance (above 92\%) while keeping user identification accuracy below 7\%.

\subsection{Deep Belief Network (DBN)}
\begin{figure}[h!]
\centering
  \includegraphics[width= 0.9\linewidth]{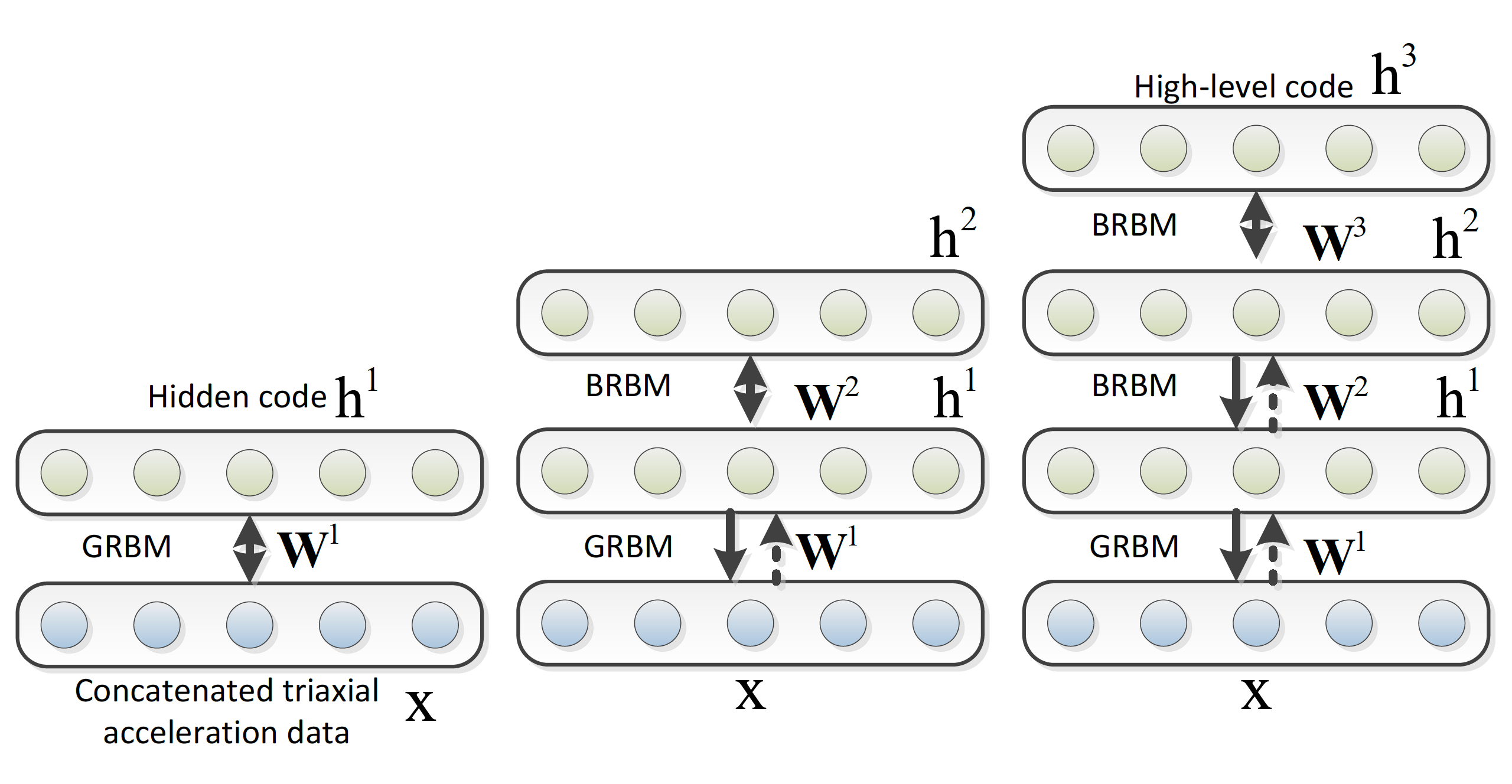}
  \caption{The greedy layer-wise training of DBNs. The first level is trained on triaxial acceleration data. Then, more RBMs are repeatedly stacked to form a deep activity recognition model~\cite{DBN4}.
  }
  \label{fig:fig_dbn}
\end{figure}

A DBN, as illustrated in Figure \ref{fig:fig_dbn}, is formed by stacking multiple simple unsupervised networks, where the hidden layer of the preceding network serves as the visible layer for the next. The representation of each sub-network is generally the restricted Boltzmann machine (RBM), an undirected generative energy-based model with a ``visible'' input layer, a hidden layer, and intra-layer connections in between. The DBN typically has connections between the layers but not between units within each layer. This structure leads to a fast and layer-wise unsupervised training procedure, where contrastive divergence (a training technique to approximate the relationship between a network's weights and its error) is applied to every pair of layers in the DBN architecture sequentially, starting from the ``lowest'' pair.
\vspace{-12pt}

The observation that DBNs can be trained greedily led to one of the first effective deep learning algorithms~\cite{DBN0}. There are many attractive implementations and uses of DBNs in real-life applications such as drug discovery~\cite{DBN8}, natural language understanding~\cite{DBN9}, fault diagnosis~\cite{DBN10}, etc. There are also many attempts to perform HAR with DBNs. In early exploratory work back in 2011~\cite{DBN1}, a five-layer DBN is trained with the input acceleration data collected from mobile phones.
The accuracy improvement ranges from 1\% to 32\% when compared to traditional ML methods with manually extracted features. 

In later works, DBN is applied to publicly available datasets~\cite{DBN2, DBN4, DBN7,mahmoodzadeh2021human}. In~\cite{DBN2}, two five-layer DBNs with different structures are applied to the Opportunity dataset~\cite{Opportunity}, USC-HAD dataset~\cite{USC-HAD}, and DSA dataset~\cite{DSA}, and the results demonstrate improved accuracy for HAR over traditional ML methods for all the three datasets. Specifically, the accuracy for the Opportunity, USC-HAD, and DSA datasets are 82.3\% (1.6\% improvement over traditional methods), 99.2\% (13.9\% improvement), and 99.1\% (15.9\% improvement), respectively. 
In addition, Alsheikh et al.~\cite{DBN4} tested the activity recognition performance of DBNs using different parameter settings. Instead of using the raw acceleration data similar to~\cite{DBN1}, they used spectrogram signals of the triaxial accelerometer data to train the deep activity recognition models. They found that deep models with more layers outperform the shallow models, and the topology of layers having more neurons than the input layer is shown to be more advantageous, which indicates overcompete representation is essential for learning deep models. The accuracy of the tuned DBN was 98.23\%, 91.5\%, and 89.38\% on the WISDM~\cite{WISDM}, Daphnet~\cite{Daphnet}, and Skoda~\cite{Skoda} benchmark datasets, respectively. In~\cite{DBN7}, a RBM is used to improve upon other methods of sensor fusion, as neural networks can identify non-intuitive predictive features largely from cross-sensor correlations and thus offer a more accurate estimation. The recognition accuracy with this architecture on the Skoda dataset reached 81\%, which is around 6\% higher than the traditional classification method with the best performance (Random Forest).

In addition to taking advantage of public datasets, there are also researchers employing DBNs on human activity or health-related recognition with self-collected datasets~\cite{parkinson2015, DBN5}. In~\cite{parkinson2015}, DBNs are employed in Parkinson's disease diagnosis to explore if they can cope with the unreliable labelling that results from naturalistic recording environments. The data was collected with two tri-axial accelerometers, with one worn on each wrist of the participant. The DBNs built are two-layer RBMs, with the first layer as a Guassian-binary RBM (containing gaussian visible units) and the second layer as binary-binary (containing only binary units) (please refer to~\cite{hinton2012practical} for details). In~\cite{DBN5}, an unsupervised five-layer DBM-DNN is applied for the automatic detection of eating episodes via commercial bluetooth headsets collecting raw audio signals, and demonstrate classification improvement even in the presence of ambient noise. The accuracy of the proposed DBM-DNN approach is 94\%, which is significantly better than SVM with a 75.6\% accuracy. 

\subsection{Convolutional Neural Network (CNN)}

A CNN comprises convolutional layers that make use of the convolution operation, pooling layers, fully connected layers, and an output layer (usually Softmax layer). The convolution operation with a shared kernel enables the learning process of space invariant features. Because each filter in a convolutional layer has a defined receptive field, CNN is good at capturing local dependency, compared with a fully-connected neural network. Though each kernel in a layer covers a limited size of input neurons, by stacking multiple layers, the neurons of higher layers will cover a larger more global receptive field. The pyramid structure of CNN contributes to its capability of gathering low-level local features into high-level semantic meanings. This allows CNN to learn excellent features as shown in~\cite{cnn91}, which compares the features extracted from CNN to hand-crafted time and frequency domain features (Fast Fourier Transform and Discrete Cosine Transform).

%

CNN incorporates a pooling layer that follows each convolutional layer in most cases. A pooling layer compresses the representation it is learning and strengthens the model against noise by dropping a portion of the output to a convolutional layer. Generally, a few fully connected layers follow after a stack of convolutional and pooling layers that reduce feature dimensionality before being fed into the output layer. A softmax classifier is usually selected as the final output layer. However, as an exception, some studies explored the use of traditional classifiers as the output layer in a CNN~\cite{cnn11,cnn115}. 

%



Most CNNs use univariate or multivariate sensor data as input. Besides raw or filtered sensor data, the magnitude of 3-axis acceleration is often used as input, as shown in~\cite{cnn2017bigcomp}. Researchers have tried encoding time-series data into 2D images as input into the CNN. In~\cite{cnn35}, the Short-time Fourier transform (STFT) for time-series sensor data is calculated, and its power spectrum is used as the input to a CNN. Since time series data is generally one-dimensional, most CNNs adopt 1D-CNN kernels. Works that use frequency-domain inputs (e.g., spectrogram), which have an additional frequency dimension, will generally use 2D-CNN kernels~\cite{cnn69}. The choice of 1D-CNN kernel size normally falls in the range of 1~$\times$~3 to 1~$\times$~5 (with exceptions in~\cite{cnnhar2014,cnn11,cnn12} where kernels of size 1~$\times$~8, 2~$\times$~101, and 1~$\times$~20 \mbox{are adopted). }

To discover the relationship between the number of layers, the kernel size, and the complexity level of the tasks, we picked and summarized several typical studies in \mbox{Table \ref{tab:CNN}.}
A majority of the CNNs consist of five to nine layers~\cite{cnn2015smc,cnn2, cnn5, cnn11, cnn12, cnn17, cnn31,huang2021shallow,gao2021deep}, usually including two to three convolutional layers, two to three max-pooling layers, followed by one to two fully connected layers before feeding the feature representation into the output layer (softmax layer in most cases). \changezs{\citet{dong2019harnet} demonstrated performance improvements by leveraging both handcrafted time and frequency domain features along with features generated from a CNN, called HAR-Net, to classify six locomotion activities using accelerometer and gyroscope signals from a smartphone.} \citet{cnn66} used a shallow three-layer CNN network including a convolutional layer, a fully connected layer, and a softmax layer to perform on-device activity recognition on a resource-limited platform and shown its effectiveness and efficiency on public datasets. \citet{cnnhar2014} and \citet{cnn2017bigcomp} also used a small number of layers (four layers). 
The choice of the loss function is an important decision in training CNNs. In classification tasks, cross-entropy is most commonly used, while in regression tasks, mean squared error is most commonly used. 
\change{Most CNN models process input data by extracting and learning channel-wise features separately while \citet{huang2021shallow} first propose a shallow CNN that considers cross-channel communication.
The channels in the same layer interact with each other to obtain discriminative features of sensor data.
}
%
%
%

\begin{table}[h!]
\centering
\renewcommand\tabcolsep{4.5pt} 
\caption{\changezs{Summary of typical studies that use layer-by-layer CNN structure in HAR and their configurations. \change{We aim to present the relationship of CNN kernels, layers, and targeted problems (application and sensors)}.} Key: C---convolutional layer; P---max-pooling layer; FC---fully connected layer; S---softmax; S1---accelerometer; S2---gyroscope; S3---magnetometer; S4---EMG; S5---ECG}

\centering 
\begin{tabular}{ccccccc}
\toprule
\textbf{Study}    & \textbf{Architecture} & \textbf{Kernel Conv. }  & \textbf{Application}       &\textbf{ \# Classes} & \textbf{Sensors}  & \textbf{Dataset} \\ \midrule
{\cite{cnn2017bigcomp}} & C-P-FC-S    & \begin{tabular}[c]{@{}c@{}}1~$\times$~3, 1~$\times$~4,\\ 1~$\times$~5\end{tabular}  & locomotion activities     & 3 & S1   & Self     \\ \midrule
{\cite{gholamrezaii2021time}} & \change{C-P-C-P-S}  & \change{4~$\times$~4} & \change{locomotion activities}  & \change{6, 12}   & \change{S1} & \change{UCI, mHealth}   \\ \midrule
{\cite{cnnhar2014}}  & C-P-FC-FC-S & 1~$\times$~20 & \begin{tabular}[c]{@{}c@{}}daily activities,\\ locomotion activities\end{tabular}                                                   & -                          & -& \begin{tabular}[c]{@{}c@{}c@{}}Skoda, \\ Opportunity, \\ Actitracker\end{tabular}           \\ \midrule
{\cite{xu2021human}} & \change{C-P-C-P-FC-S}  & \change{5~$\times$~5} & \change{locomotion activities}  & \change{6}   & \change{S1} & \change{WISDM}   \\ \midrule
{\change{\cite{uddin2018activity}}} & \change{C-P-C-P-C-FC}   &  \change{$1\times5, 1\times9$} & \change{locomotion activities}  & \change{12}   & \change{S5} & \change{mHealth}   \\ \midrule
{\cite{cnn13}} & C-P-C-P-FC-FC-S & -               & \begin{tabular}[c]{@{}c@{}}daily activities,\\ locomotion activities\end{tabular}                    & 12      & \begin{tabular}[c]{@{}c@{}}S1, S2, S3 \\ ECG\end{tabular} & mHealth          \\ \midrule
{\cite{cnn89}} & C-P-C-P-C-P-S & 12~$\times$~2            & \begin{tabular}[c]{@{}c@{}}daily activities including \\ brush teeth, comb hair, \\ get up from bed, etc\end{tabular}               & 12                                                               & S1, S2, S3                                                                   & WHARF                                                                                \\ \midrule
{\cite{cnn2015smc}}  & C-P-C-P-C-P-S                                                                                         &   12~$\times$~2      & locomotion activities   & 8 & S1    & Self  \\ \midrule
{\cite{cnn5}}  & \begin{tabular}[c]{@{}c@{}}C-P-C-P-U-FC-S, \\ U: unification layer\end{tabular}                       & 1~$\times$~3, 1~$\times$~5        & \begin{tabular}[c]{@{}c@{}}daily activities,\\ hand gesture\end{tabular}  & \begin{tabular}[c]{@{}c@{}}18 (Opp)\\ 12 (hand)\end{tabular} & \begin{tabular}[c]{@{}c@{}}S1, S2 \\ (1 for each)\end{tabular}               & \begin{tabular}[c]{@{}c@{}}Opportunity\\ Hand Gesture\end{tabular}                   \\ \midrule
{\cite{cnn12}} & C-C-P-C-C-P-FC  & 1~$\times$~8  & hand motion classification & 10    & S4     & \begin{tabular}[c]{@{}c@{}}Rami EMG\\Dataset\end{tabular}       \\ \midrule
{\cite{cnn17}} & \begin{tabular}[c]{@{}c@{}}C-C-P-C-C-\\P-FC-FC-S \\ (one branch \\ for each sensor)\end{tabular} & 1~$\times$~5             & \begin{tabular}[c]{@{}c@{}}daily activities,\\  locomotion activities,\\  industrial ordering \\ picking recognition task\end{tabular} & \begin{tabular}[c]{@{}c@{}}18 (Opp)\\ 12 (PAMAP2)\end{tabular}   &   S1, S2, S3     & \begin{tabular}[c]{@{}@{}c@{}}Opportunity,\\PAMAP2,\\ Order Picking\end{tabular} \\ \midrule
{\cite{cnn35}} & \begin{tabular}[c]{@{}c@{}}C-P-C-P-C-P-\\ FC-FC-FC-S\end{tabular}   & \begin{tabular}[c]{@{}c@{}}1~$\times$~4, 1~$\times$~10,\\ 1~$\times$~15\end{tabular}  & locomotion activities  & 6   & S1, S2, S3 & Self   \\ \midrule
\end{tabular}

\label{tab:CNN}
\end{table}

The number of sensors used in a HAR study can vary from a single one to as many as 23~\cite{Opportunity}. In~\cite{cnn2015smc}, a single accelerometer is used to collect data from three locations on the body: Cloth pocket, trouser pocket and waist. The authors collect data on 100 subjects, including eight activities such as falling, running, jumping, walking, walking quickly, step walking, walking upstairs, and walking downstairs. 
Moreover, HAR applications can involve multiple sensors of different types. 
To account for all these different types of sensors and activities, \citet{cnn49} proposed a multi-branch CNN architecture. 
A multi-branch design adopts a parallel structure that trains separate kernels for each IMU sensor and concatenates the output of branches at a late stage, after which one or more fully connected layers are applied on the flattened feature representation before feeding into the final output layer. For instance, a CNN-IMU architecture contains {m} parallel branches, one per IMU. Each branch contains seven layers, then the outputs of each branch are concatenated and fed into a fully connected and a softmax output layer.
\change{\citet{GAO2021107728} has introduced a novel dual attention module including channel and temporal attention to improving the representation learning ability of a CNN model. Their method has outperformed regular CNN considerably on a number of public datasets such as PAMAP2~\cite{PAMAP2}, WISDM~\cite{WISDM}, UNIMIB SHAR~\cite{SHAR}, and Opportunity~\cite{Opportunity}.}

Another advantage of DL is that the features learned in one domain can be easily generalized or transferred to other domains. The same human activities performed by different individuals can have drastically different sensor readings. To address this challenge, \citet{cnn35} adapted their activity recognition to each individual by adding a few hidden layers and customizing the weights using a small amount of individual data. They were able to show a 3\% improvement in recognition performance.

\subsection{Recurrent Neural Network (RNN)}
Initially, the idea of using temporal information was proposed in 1991~\cite{rnn4} to recognize a finger alphabet consisting of 42 symbols and in 1995~\cite{rnn21} to classify 66 different hand shapes with about 98\% accuracy. 
Since then, the recurrent neural network (RNN) with time series as input has been widely applied to classify human activities or estimate hand gestures~\cite{rnn35, rnn6, rnn91, rnn95, rnn96, rnn104, rnn71,alessandrini2021recurrent}. 

\begin{figure}[h!]
\centering
  \includegraphics[width=.8\linewidth]{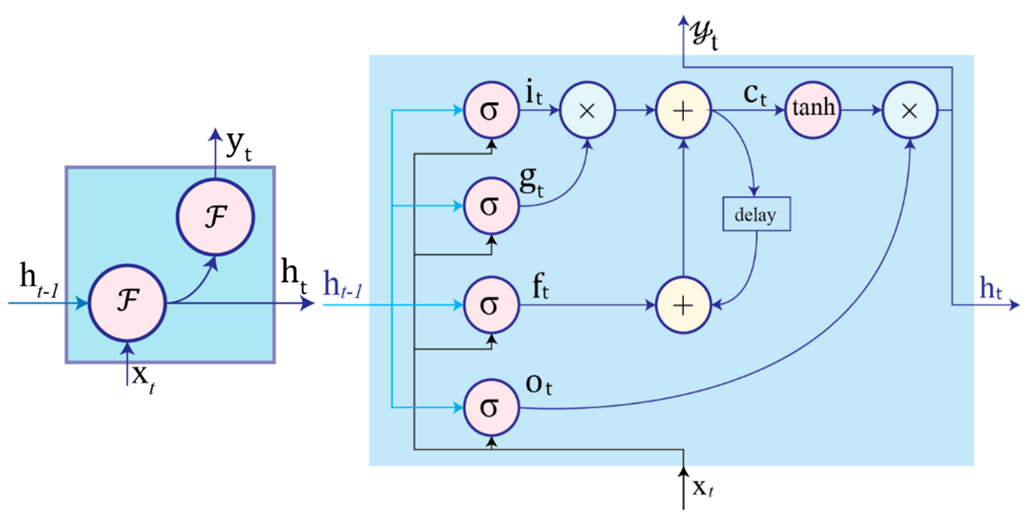}
  \caption{Schematic diagram of an RNN node and LSTM cell~\cite{rnn3}. Left: RNN node where $h_{t-1}$ is the previous hidden state, $x_t$ is the current input sample data, $h_t$ is the current hidden state, $y_t$ is the current output, and $\mathscrbf{F}$ is the activation function. Right: LSTM cell with internal recurrence $c_t$ and outer recurrence $h_t$.}%
  \label{fig:RNN_LSTM}
\end{figure}

Unlike feed-forward neural networks, an RNN processes the input data in a recurrent behavior. Equivalent to a directed graph, RNN exhibits dynamic behaviors and possesses the capability of modelling temporal and sequential relationships due to a hidden layer with recurrent connections. A typical structure for an RNN is shown in Figure \ref{fig:RNN_LSTM} with the current input, $x_t$, and previous hidden state, $h_{t-1}$. The network generates the current hidden state, $h_t$, and output, $y_t$, is as follows:

\begin{equation}
\begin{array}{c}
h_{t}=\mathscrbf{F}\left(W_{h} h_{t-1} + U_{h} x_{t} + b_{h}\right) \\
y_{t}=\mathscrbf{F}\left(W_{y} h_{t} + b_{y}\right)
\end{array}
\end{equation}
where $W_{h}$, $U_{h}$, and $W_{y}$ are the weights for the hidden-to-hidden recurrent connection, input-to-hidden connection, and hidden-to-output connection, respectively. $b_h$ and $b_y$ are bias terms for the hidden and output states, respectively. Furthermore, each node is associated with an element-wise non-linearity function as an activation function $\mathscrbf{F}$ such as the sigmoid, hyperbolic tangent (tanh), or rectified linear unit (ReLU).

In addition, many researchers have undertaken extensive work to improve the performance of RNN models in the context of human activity recognition and have proposed various models based on RNNs, including Independently RNN (IndRNN)~\cite{rnn5}, Continuous Time RNN (CTRNN)~\cite{rnn17}, Personalized RNN (PerRNN)~\cite{rnn64},  Colliding Bodies Optimization RNN (CBO-RNN)~\cite{rnn58}. 
Unlike previous models with one-dimension time-series input, \citet{rnn31} builds a CNN + RNN model with stacked multisensor data in each channel for fusion before feeding into the CNN layer. 
\citet{rnn74} uses an RNN to address the domain adaptation problem caused by intra-session, sensor placement, and intra-subject variances.

HAR improves with longer context information and longer temporal intervals. However, this may result in vanishing or exploding gradient problems while backpropagating gradients~\cite{bengiogradient}.
In an effort to address these challenges, long short-term memory (LSTM)-based RNNs~\cite{LSTM}, and Gated Recurrent Units (GRUs)~\cite{GRU} are introduced to model temporal sequences and their broad dependencies. 
The GRU introduces a reset and update gate to control the flow of inputs to a cell~\cite{rnn23,rnn28, rnn61, rnn48, rnn113}.
The LSTM has been shown capable of memorizing and modelling the long-term dependency in data. Therefore, LSTMs have taken a dominant role in time-series and textual data analysis. It has made substantial contributions to human activity recognition, speech recognition, handwriting recognition, natural language processing, video analysis, etc. 
As illustrated in \mbox{Figure \ref{fig:RNN_LSTM}~\cite{rnn3}}, a LSTM cell is composed of: (1) input gate, $i_t$, for controlling flow of new information; (2)~forget gate, $f_t$, setting whether to forget content according to internal state; (3) output gate, $o_t$, controlling output information flow; (4) input modulation gate, $g_t$, as main input; (5) internal state, $c_t$, dictates cell internal recurrence; (6) hidden state, $h_t$, contains information from samples encountered within the context window previously. The relationship between these variables are listed as Equation \eqref{equ:lstm}~\cite{rnn3}. 

\begin{equation}
\left\{ 
\begin{array}{c}
i_{t}=\sigma\left(b_{i} + U_{i} x_{t} + W_{i} h_{t-1}\right) \\
f_{t}=\sigma\left(b_{f} + U_{f} x_{t} + W_{f} x_{t-1}\right) \\
o_{t}=\sigma\left(b_{o} + U_{o} x_{t} + W_{o} h_{t-1}\right) \\
g_{t}=\sigma\left(b_{g} + U_{g} x_{t} + W_{g} h_{t-1}\right) \\
c_{t}=f_{t} c_{t-1} + g_{t} i_{t} \\
h_{t}=\tanh \left(c_{t}\right) o_{t}
\end{array}
\right.
\label{equ:lstm}
\end{equation}

\vspace{-6pt}

\begin{figure}[h!]
\centering
  \includegraphics[width=0.85\linewidth]{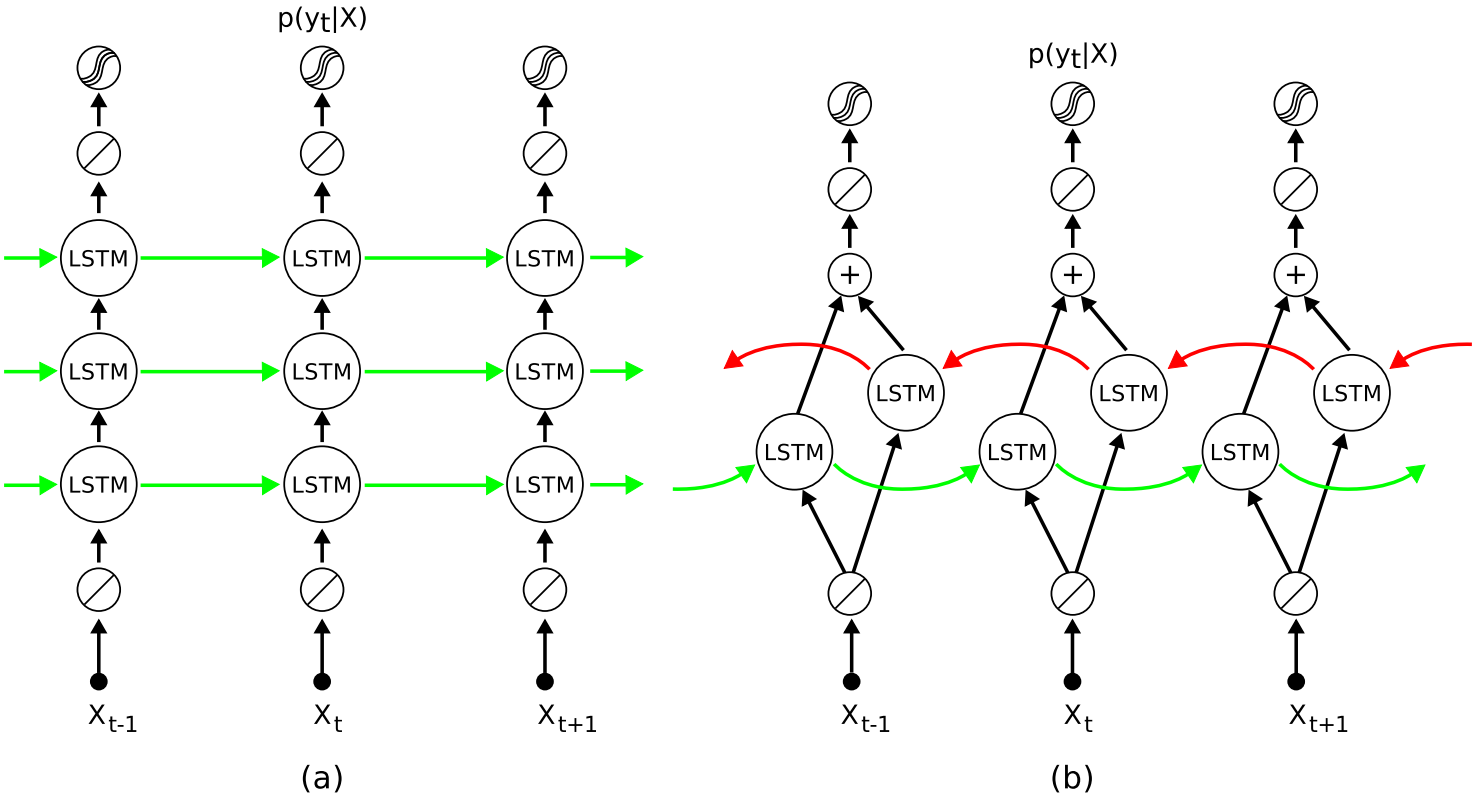}
  \caption{The structure of LSTM and bi-directional LSTM model~\cite{hammerla2016deep}. (\textbf{a}). LSTM network hidden layers containing LSTM cells and a final softmax layer at the top. (\textbf{b}) bi-directional LSTM network with two parallel tracks in both future (green) and past (red) directions.}
  \label{fig:lstm}
\end{figure}

As shown in Figure \ref{fig:lstm}, the input time series data is segmented into windows and fed into the LSTM model. For each time step, the model computes class prediction scores, which are then merged via late-fusion and used to calculate class membership probabilities through the softmax layer. 
Previous studies have shown that LSTMs have high performance in wearable HAR~\cite{rnn61, rnn3, rnn76}.
Researchers in~\cite{hammerla2016deep} rigorously examine the impact of hyperparameters in LSTM with the fANOVA framework across three representative datasets, containing movement data captured by wearable sensors. 
The authors assessed thousands of settings with random hyperparameters and provided guidelines for practitioners seeking to apply deep learning to their own problem scenarios~\cite{hammerla2016deep}. Bidirectional LSTMs, having both past and future recurrent connections, were used in~\cite{rnn30, rnn80} to classify activities.

\vspace{-6pt}

Researchers have also explored other architectures involving LSTMs to improve benchmarks on HAR datasets. Residual networks possess the advantage that they are much easier to train as the addition operator enables gradients to pass through more directly. Residual connections do not impede gradients and could help to refine the output of layers. For example,~\cite{rnn48} proposes a harmonic loss function and~\cite{rnn16} combines LSTM with batch normalization to achieve 92\% accuracy with raw accelerometer and gyroscope data.
Ref.~\cite{lstm7} proposes a hybrid CNN and LSTM model (DeepConvLSTM) for activity recognition using multimodal wearable sensor data. DeepConvLSTM performed significantly better in distinguishing closely-related activities, such as ``Open/Close Door'' and ``Open/Close Drawer''.
Moreover, Multitask LSTM is developed in~\cite{lstm38} to first extract features with shared weight, and then classify activities and estimate intensity in separate branches. Qin \etal proposed a deep-learning algorithm that combines CNN and LSTM networks~\cite{lstm114}. They achieved 98.1\% accuracy on the SHL transportation mode classification dataset with CNN-extracted and hand-crafted features as input. Similarly, other researchers~\cite{lstm25, lstm26, lstm12, lstm133, lstm130, lstm121, lstm100, lstm64,saha2021deep} have also developed the CNN-LSTM model in various application scenarios by taking advantage of the feature extraction ability of CNN and the time-series data reasoning ability of LSTM. 
\change{Interestingly, utilizing CNN and LSTM combined model, researchers in~\cite{saha2021deep} attempt to eliminate sampling rate variability, missing data, and misaligned data timestamps with data augmentation when using multiple on-body sensors. Researchers in~\cite{mekruksavanich2021placement} explored the placement effect of motion sensors and discovered that the chest position is ideal for physical activity identification.}

\textls[-15]{Raw IMU and EMG time series data are commonly used as inputs to RNNs~\cite{rnn1, rnn26, rnn41, rnn55, rnn74, rnn121}}. A number of major datasets used to train and evaluate RNN models have been created, including the Sussex-Huawei Locomotion-Transportation (SHL)~\cite{rnn5, rnn28}, PAMAP2~\cite{rnn27, rnn31} and Opporunity\cite{rnn76}. In addition to raw time series data~\cite{rnn61}, 
Besides raw time series data, custom features are also commonly used as inputs to RNNs.~\cite{rnn23} showed that training an RNN with raw data and with simple custom features yielded similar performance for gesture recognition (96.89\% vs 93.38\%).

However, long time series may have many sources of noise and irrelevant information. 
The concept of attention mechanism was proposed in the domain of neural machine translation to address the problem of RNNs being unable to remember long-term relationships. The attention module mimics human visual attention to building direct mappings between the words/phrases that represent the same meaning in two languages. It eliminates the interference from unrelated parts of the input when predicting the output. This is similar to what we as humans perform when we translate a sentence or see a picture for the first time; we tend to focus on the most prominent and central parts of the picture. An RNN encoder attention module is centred around a vector of importance weights. The weight vector is computed with a trainable feedforward network and is combined with RNN outputs at all the time steps through the dot product. The feedforward network takes all the RNN immediate outputs as input to learn the weights for each time step. 
\cite{rnn113} utilizes attention in combination with a 1D CNN Gated Recurrent Units (GRUs), achieving HAR performances of 96.5\% ± 1.0\%, 93.1\% ± 2.2\%, and 89.3\% ± 1.3\% on Heterogeneous~\cite{Stisen}, Skoda~\cite{Skoda}, and PAMAP2~\cite{PAMAP2} datasets, respectively. 
\cite{rnn27} applies temporal attention and sensor attention into LSTM to improve the overall activity recognition accuracy by adaptively focusing on important time windows and sensor modalities.

In recent years, block-based modularized DL networks have been gaining traction. Some examples are GoogLeNet with an Inception module and Resnet with residual blocks. The HAR community is also actively exploring the application of block-based networks. In~\cite{innohar}, the authors have used GoogLeNet’s Inception module combined with a GRU layer to build a HAR model. The proposed model was showed performance improvements on three public datasets (Opportunity, PAMAP2 and Smartphones datasets). \change{\citet{qian2019novel} developed the model with $SMM_{AR}$ in a statistical module to learn all orders of moments statistics as features, LSTM in a spatial module to learn correlations among sensors placements, and LSTM + CNN in a temporal module to learn temporal sequence dependencies along the \mbox{time scale.}
}


\subsection{Deep Reinforcement Learning (DRL)}

\begin{figure}[h!]
\centering
  \includegraphics[width=0.7\columnwidth]{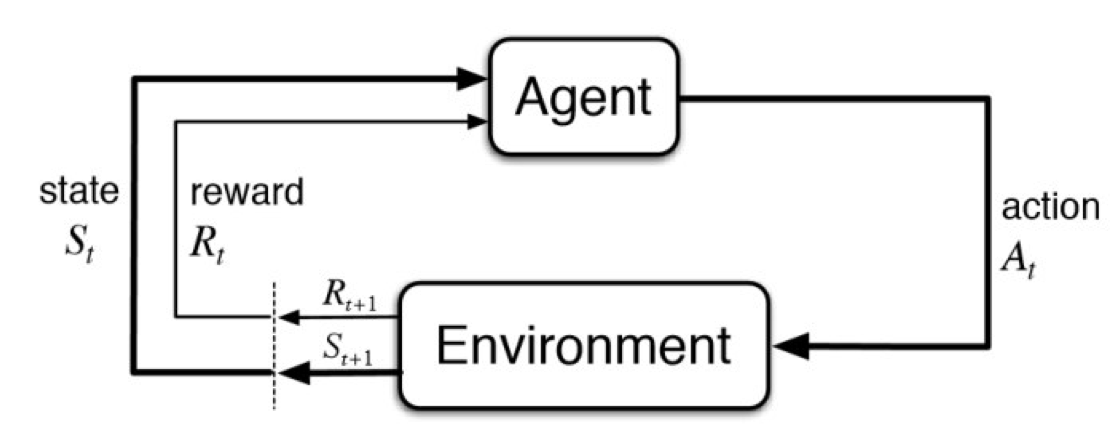}
  \caption{A typical structure of a reinforcement learning network~\cite{rl-img}.}
  \label{fig:rl}
\end{figure}

AE, DBN, CNN\changezs{, and RNN} fall within the realm of supervised or unsupervised learning. Reinforcement learning is another paradigm where an agent attempts to learn optimal policies for making decisions in an environment. At each time step, the {agent} takes an {action} and then receives a {reward} from the {environment}. The {state} of the environment accordingly changes with the action made by the agent. The goal of the agent is to learn the (near) optimal policy (or probability of action, state pairs) through the interaction with the environment in order to maximize a cumulative long-term reward. The two entities---agent and environment---and the three key elements---action, state and reward---collectively form the paradigm of RL. The structure of RL is shown in Figure~\ref{fig:rl}. 

\vspace{-6pt}


In the domain of HAR,~\cite{rl11} uses DRL to predict arm movements with 98.33\% accuracy.
Ref. \cite{rl2}~developed a reinforcement learning model for imitating the walking pattern of a lower-limb amputee on a musculoskeletal model. The system showed 98.02\% locomotion mode recognition accuracy. Having a high locomotion recognition accuracy is critical because it helps lower-limb amputees prevent secondary impairments during rehabilitation.
In~\cite{bhat2018online}, Bhat et al. propose a HAR online learning framework that takes advantage of reinforcement learning utilizing a policy gradient algorithm for faster convergence achieving 97.7\% in recognizing six activities.

\subsection{Generative Adversarial Network (GAN)}
\label{sec:gan}

%

\begin{figure}[h!]	

\includegraphics[width=1\linewidth]{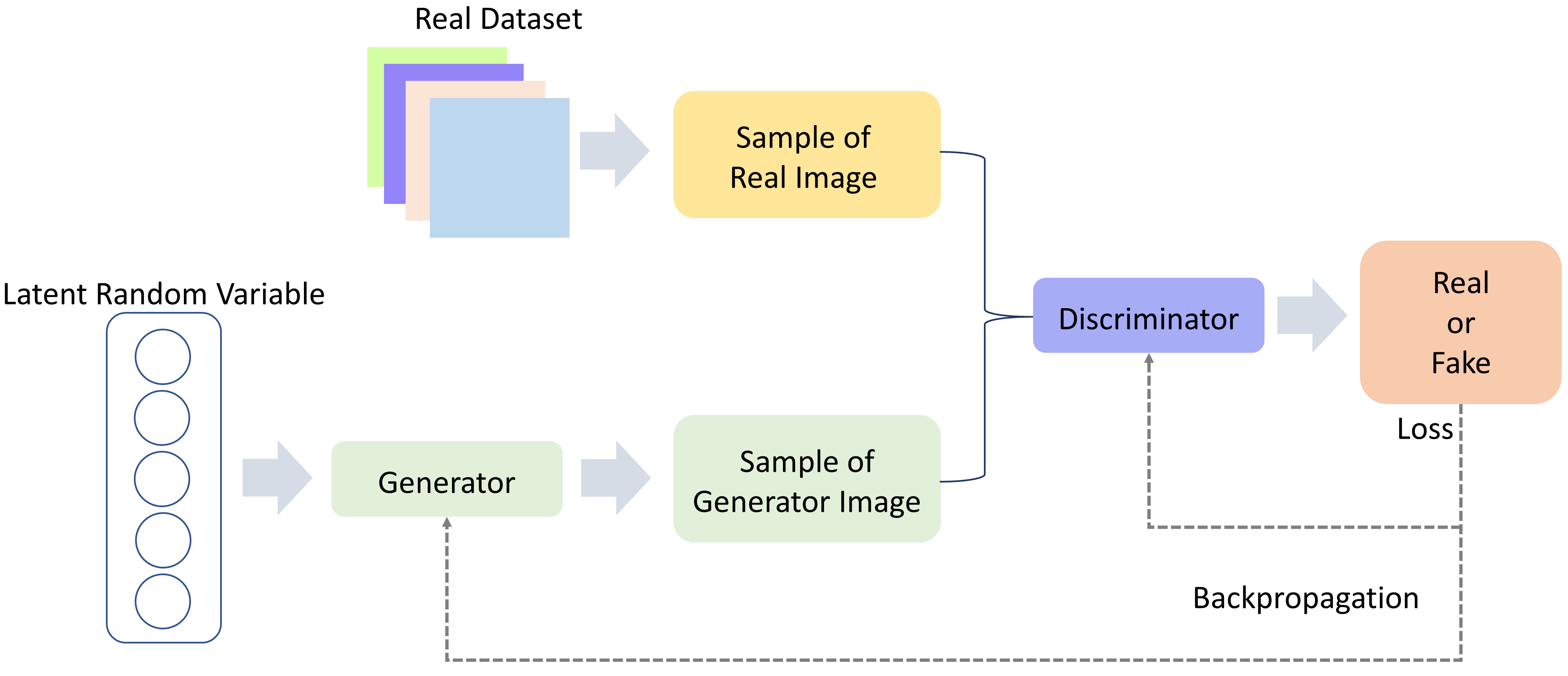}
\caption{The structure of generative adversarial network.}
\label{fig:gan}
\end{figure}

Originally proposed to generate credible fake images that resemble the images in the training set, 
GAN is a type of deep generative model, which is able to create new samples after learning from real data~\cite{goodfellow2014generative}. 
It comprises two networks, the generator ($\mathbf{G}$) and the discriminator ($\mathbf{D}$), competing against each other in a zero-sum game framework as shown in Figure \ref{fig:gan}. During the training phase, the generator takes as input a random vector $\mathbf{z}$ and transforms $\mathbf{z}\in \mathbb{R^\mathbf{n}}$ to plausible synthetic samples $\mathbf{\hat{x}}$ to challenge the discriminator to differentiate between original samples $\mathbf{x}$ and fake samples $\mathbf{\hat{x}}$. In this process, the generator strives to make the output probability $\mathbf{D(G(z))}$ approach one, in contrast with the discriminator, which tries to make the function's output probability as close to zero as possible. The two adversarial rivals are optimized by finding the Nash equilibrium of the game in a zero-sum game setting, which means the adversarial rivals' gains would be maintained regardless of what strategies are selected. However, it is not theoretically guaranteed that GAN zero-sum games reach Nash Equilibria~\cite{farnia2020gans}.

GAN model has shown remarkable performance in generating synthetic data with high quality and rich details~\cite{springenberg2015unsupervised, odena2016semi}. In the field of HAR, GAN has been applied as a semi-supervised learning approach to deal with unlabeled or partially labelled data for improving performance by learning representations from the unlabeled data, which later will be utilized by the network to generalize to the unseen data distribution~\cite{shi2021gan}. 
Afterwards, GAN has shown the ability to generate balanced and realistic synthetic sensor data. 
\citet{SensoryGANs} utilized GANs with a customized network to generate synthetic data from the public HAR dataset HASC2010corpus~\cite{HASC2011corpus}. 
Similarly, \citet{gan78} assessed synthetic data with CNN or LSTM models as a generator. In two public datasets, Sussex-Huawei Locomotion (SHL) and Smoking Activity Dataset (SAD), the discriminator was built with CNN layers, and the results demonstrated synthetic data with high quality and diversity with two public datasets. Moreover, by oversampling and adding synthetic sensor data into the training, researchers augmented and alleviated the originally imbalanced training set to achieve better performance. 
In~\cite{gan26, gan69}, they generated verisimilar data of different activities, and \citet{gan27} used the Boulic kinematic model, which aims to capture the three-dimensional positioning trend to synthesize personified walking data.
\change{Due to the ability to generate new data, GAN has been widely applied in transfer learning in HAR to help with the dramatic performance drop when the pre-trained model are tested against unseen data from new users.
In transfer learning techniques, the learned knowledge from the source domain (subject) is transferred to the target domain to decrease the lack of performance of the models within the target domain. 
Moreover,~\cite{soleimani2021cross} is an attempt that utilized GAN to perform cross-subject transfer learning for HAR since collecting data for each new user was infeasible. 
With the same idea, cross-subject transfer learning based on GAN outperformed those without GAN on Opportunity benchmark dataset in~\cite{soleimani2021cross} and outperformed unsupervised learning on UCI and USC-HAD dataset~\cite{abedin2021guidedgan}. Even more, transfer learning under conditions of cross-body, cross-user, and cross-sensor has been demonstrated superior performance in~\cite{sanabria2021contrasgan}.}

However, much more effort is needed in generating verisimilar data to alleviate the burden and cost of collecting sufficient user data.
Additionally, it is typically challenging to obtain well-trained GAN models owing to the wide variability in amplitude, frequency, and period of the signals obtained from different types of activities.

\subsection{\change{Hybrid Models}}
\label{sec:hybrid}

\change{
As an advancement of machine learning models, researchers take advantage of different methods and propose hybrid models.
The combination of CNN and LSTM endows the model capability of extracting local features as well as long-term dependencies in sequential data, especially for HAR time series data. For example, \citet{challa2021multibranch} proposed a hybrid of CNN and bidirectional long short-term memory (BiLSTM). The accuracy on UCI-HAR, WISDM~\cite{WISDM}, and PAMAP2~\cite{PAMAP2} datasets achieved 96.37\%, 96.05\%, and 94.29\%, respectively. \citet{dua2021multi} proposed a model with CNN combined with GRU and obtained an accuracy of 96.20\%, 97.21\%, and 95.27\% on UCI-HAR, WISDM \cite{WISDM}, and PAMAP2~\cite{PAMAP2} datasets, respectively.
In order to have a straightforward view of the functionality of hybrid models, we list several papers with CNN only, LSTM only, CNN + GRU, and CNN + LSTM in \mbox{Tables \ref{tab:UCI-HAR_hybrid} and  \ref{tab:PAMAP2_hybrid}}.
In addition, \citet{RL1} proposed to combine reinforcement learning and LSTM model to improve the adaptability of different kinds of sensors, including EEG (EID dataset), RFID (RSSI dataset)~\cite{yao2017compressive}, and wearable IMU (PAMAP2 dataset)~\cite{PAMAP2}.
\mbox{Ref. \cite{zhang2019know}} employed CNN for feature extraction and a reinforced selective attention model to automatically choose the most characteristic information from multiple channels.
}

\begin{table}[h!]
\centering
\renewcommand\tabcolsep{28.2pt} 
    
     \caption{\change{Comparison of models on UCI-HAR dataset.}}
    \begin{tabular}{lcc}
    \toprule
    \textbf{Model} & \textbf{F1-Score (\%) }&\textbf{ Accuracy (\%) } \\
    \midrule
    CNN~\cite{wan2020deep} &	92.93 &	92.71 \\
    Res-LSTM~\cite{zhao2018deep} & 91.50 & 91.60  \\
    Stacked-LSTM~\cite{ullah2019stacked} &	-- &	93.13 \\
    CNN-LSTM~\cite{lstm130} & -- &	92.13 \\
    Bidir-LSTM~\cite{hernandez2019human}  &	-- &	92.67 \\
    Residual-BiLSTM~\cite{zhao2018deep} &	93.5 &	93.6 \\
    LSTM-CNN~\cite{lstm25}  &	-- &	95.78 \\ 
    CNN‑GRU~\cite{dua2021multi} &	-- & 96.20 \\
    CNN‑GRU~\cite{challa2021multibranch} &	94.54 & 94.58 \\
    CNN-LSTM~\cite{challa2021multibranch} & 94.76 & 94.80 \\
    CNN‑BiLSTM~\cite{challa2021multibranch} &	96.31 &	96.37 \\
    \bottomrule
    \end{tabular}
    \label{tab:UCI-HAR_hybrid}

\end{table}
\vspace{-9pt}
\begin{table}[h!]
\centering
    \renewcommand\tabcolsep{31.5pt} 
     \caption{\change{Comparison of models on PAMAP2 dataset.}}
    \begin{tabular}{lcc}
    \toprule
    \textbf{Model} & \textbf{F1-Score (\%)} & \textbf{Accuracy (\%)}  \\
    \midrule
    CNN\cite{wan2020deep} & 91.16 & 91.00 \\
    BiLSTM~\cite{hernandez2019human} & 89.40 & 89.52   \\
    LSTM-F~\cite{hammerla2016deep}  &	 92.90  & -- \\
    COND-CNN~\cite{cheng2020real} &	--  &94.01 \\
    CNN‑GRU~\cite{dua2021multi} & 	-- & 95.27 \\
    CNN‑GRU~\cite{challa2021multibranch} &	93.16 & 93.20 \\
    CNN-LSTM~\cite{challa2021multibranch} & 92.77 & 92.81 \\
    CNN‑BiLSTM~\cite{challa2021multibranch} &	94.27  & 94.29 \\
    \bottomrule
    \end{tabular}    
    \label{tab:PAMAP2_hybrid}

\end{table}




%

\subsection{Summary and Selection of Suitable Methods }
\label{sec:summary}

\change{
Since the last decade, DL methods have gradually dominated a number of artificial intelligence areas, including sensor-based human activity recognition, due to its automatic feature extraction capability, strong expressive power, and the high performance rendered. 
When a sufficient amount of data are available, we are becoming prone to turn to DL methods. 
With all these types of available DL approaches discussed above, we need to get a full understanding of the pros and cons of these approaches in order to select the appropriate approach wisely. To this end, we briefly analyze the characteristics of each approach and attempt to give readers high-level guidance on how to choose the DL approach according to the needs and requirements.} 

\change{The most salient characteristic of auto-encoder is that it does not require any annotation. Therefore, it is widely adopted in the paradigm of unsupervised learning. Due to its exceptional capability in dimension reduction and noise suppression, it is often leveraged to extract low-dimensional feature representation from raw input. However, auto-encoders may not necessarily learn the correct and relevant characteristics of the problem at hand. There is also generally little insight that can be gained for sensor-based auto-encoders, making it difficult to know which parameters to adjust during training.
Deep belief networks are a generative model generally used for solving unsupervised tasks by learning low-dimensional features. Today, DBNs have been less often chosen compared with other DL approaches and are rarely used due to the tedious training process and increased training difficulty with DBN when the network goes deeper~\cite{Goodfellow-et-al-2016}.
}

\change{ 
CNN architecture is powerful to extract hierarchical features owing to its layer-by-layer hierarchical structure. 
When compared with other approaches like RNN and GAN, CNN is relatively easy to implement. 
Besides, as one of the most studied DL approaches in image processing and computer vision,  there is a large range of CNN variants existing that we can choose from to transfer to sensor-based HAR applications. 
When sensor data are represented as two-dimensional input, we can directly start with pre-trained models on a large image dataset (e.g., ImageNet) to fasten the convergence process and achieve better performance. 
Therefore, adapting the CNN approach enjoys a higher degree of flexibility in the available network architecture (e.g., GoogLeNet, MobileNet, ResNet, etc) than other DL approaches.
However, CNN architecture has the requirement of fixed-sized input, in contrast to RNN, which accepts flexible input size.
%
In addition, compared with unsupervised learning methods such as auto-encoder and DBN, a large number of annotated data are required, which usually demands expensive labelling resources and human effort to prepare the dataset. 
The biggest advantage of RNN and LSTMs is that they can model time series data (nearly all sensor data) and temporal relationships very well. Additionally, RNN and LSTMs can accept flexible input data size. The factors that prevent RNN and LSTMs from becoming the de facto method in DL-based HAR is that they are difficult to train in multiple aspects. They require a long training time and are very susceptible to diminishing/exploding gradients. It is also difficult to train them to efficiently model long time series.
}

\change{
GAN, as a generative model, can be used as a data augmentation method. Because it has a strong expressive capability to learn and imitate the latent data distribution of the targeted data, it outperforms traditional data augmentation methods~\cite{parkinson_augmentation}. 
Owing to its inherent data augmentation ability, GAN has the advantage of alleviating data demands at the beginning. 
However, GAN is often considered as hard to train because it alternatively trains a generator and a discriminator. Many variants of GAN and special training techniques have been proposed to tackle the converging issue~\cite{arjovsky2017wasserstein, gulrajani2017improved, che2016mode}.
}

\change{
Reinforcement learning is a relatively new area that is being explored for select areas in HAR, such as modelling muscle and arm movements~\cite{rl11, rl2}. Reinforcement learning is a type of unsupervised learning because it does not require explicit labels. Additionally, due to its online nature, reinforcement learning agents can be trained online while deployed in a real system. However, reinforcement learning agents are often difficult and time-consuming to train. Additionally, in the realm of DL-based HAR, the reward of the agent has to be given by a human, as in the case of~\cite{rl11,bhat2018online}. In other words, even though people do not have to give explicit labels, humans are still required to provide something akin to a label (the reward) to train the agent.
}

\change{
When starting to choose a DL approach, we have a list of factors to consider, including the complexity of the target problem, dataset size, the availability and size of annotation, data quality, available computing resource, as well as the requirement of training time. 
%
Firstly, we have to evaluate and examine the problem complexity to decide upon promising venues of machine learning methods. For example, if the problem is simple enough to resolve with the provided sensor modality, it's very likely that manual feature engineering and traditional machine learning method can provide satisfying results thus no DL method is needed. 
%
Secondly, before we choose the routine of DL, we would like to make sure the dataset size is sufficient to support a DL method. 
The lack of a sufficiently large corpus of labelled high-quality data is a major reason why DL methods cannot produce an expected result.
Normally, when training a DL model with a limited dataset size, the model will be prone to overfitting, and the generalizability will be sacrificed, thus using a very deep network may not be a good choice. One option is to go for a shallow neural network or a traditional ML approach. Another option is to utilize specific algorithms to make the most out of the data. To be specific, data augmentation methods such as GAN can be readily implemented.
%
Thirdly, another determining factor is the availability and size of annotation. When there is a large corpus of unlabeled sensor data at hand, a semi-supervised learning scheme is a promising direction one could consider, which will be discussed later in this work. 
%
Besides the availability of sensor data, the data quality also influences the network design. If the sensor is vulnerable to environmental noise, inducing a small SNR, some type of denoising structure (e.g., denoising auto-encoder) and increasing depth of the model can be considered to increase the noise-resiliency of the DL model.
%
At last, a full evaluation of available computing resources and expected model training time cannot be more important for developers and researchers to choose a suitable DL approach.
}

\section{Challenges and Opportunities} \label{sec:challenges}

Though HAR has seen rapid growth, there are still a number of challenges that, if addressed, could further improve the status quo, leading to increased adoption of novel HAR techniques in existing and future wearables. In this section, we discuss these challenges and opportunities in HAR. 
Note that the issues discussed here are applicable to general HAR, not only DL-based HAR. We look to discuss and analyze the following \change{four questions under our research question $\mathsf{Q3}$ (challenges and opportunities)}, which overlap with the \change{four major constituents} of machine learning.

\begin{itemize}
    \item $\mathsf{Q3.1:}$ What are the challenges in data acquisition? How do we resolve them? 
    \item $\mathsf{Q3.2:}$ What are the challenges in label acquisition? What are the current methods? 
    \item $\mathsf{Q3.3:}$ What are the challenges in modeling? What are potential solutions?
    \item $\mathsf{Q3.4:}$ \change{What are the challenges in model deployment? What are potential opportunities?}
\end{itemize}

\subsection{Challenges in Data Acquisition}

Data is the cornerstone of artificial intelligence. \minorrevision{Models only perform as well as the quality of the training data. To build generalizable models, careful attention should be paid to data collection, ensuring the participants are representative of the population of interest. Moreover, determining a sufficient training dataset size is important in HAR. Currently, there is no well-defined method for determining the sample size of training data. However, showing the convergence of the error rate as a function of training data size is one approach shown by Yang \etal~\cite{voice_in_ear}. Acquiring a massive amount of high-quality data at a low cost is critical in every domain. 
In HAR, collecting raw data is labor-intensive considering a large number of different wearables. Therefore, proposing and developing innovative approaches to augmenting data with high quality is imperative for the growth of \mbox{HAR research.}} 

\subsubsection{The Need for More Data}
\label{subsubsec:the_need_for_more_data}

\minorrevision{Data collection requires a considerable amount of effort in HAR. Particularly when researchers propose their original hardware, it is inevitable to collect data on users. 
Data augmentation is commonly used to generate synthetic training data when there is a data shortage. 
Synthetic noise is applied to real data to obtain new training samples. In general, using the dataset augmented with synthetic training samples yields higher classification accuracy when compared to using the original dataset \cite{parkinson_augmentation, lstm_augmentation, resnet_augmentation, shoes}}. 
 Giorgi \etal augmented their dataset by varying each signal sample with translation drawn from a small uniform distribution and showed improvements in accuracy using this augmented dataset~\cite{shoes}. 
\minorrevision{
\citet{resnet_augmentation} utilized Dynamic Time Warping to augment data and tested on UCR archive~\cite{UCRArchive}.
Deep learning methods are also used to augment the datasets to improve performance~\cite{T-CGAN, SensoryGANs, SenseGen}. 
\citet{SenseGen} and \citet{SensoryGANs} employed GAN to synthesize sensor data using existing sensor data. 
\citet{T-CGAN} designed a conditional GAN-based framework to generate new irregularly-sampled time series to augment unbalanced data sets.
Several works extracted 3D motion information from videos and transferred the knowledge to synthesize virtual on-body IMU sensor data~\cite{IMUTube,vid2finger}. In this way, they realized cross-modal IMU sensor data generation using traditional computer vision and graphics methods.
\textbf{Opportunity:} 
We have listed some of the most recent works focusing on cross-modal sensor data synthesis. However, few researchers (if any) used a deep generative model to build a video-sensor multi-modal system. If we take a broader view, many works are using cross-modal deep generative models (such as GAN) in data synthesis, such as from video to audio~\cite{visual2sound}, from text to image and vice versa~\cite{imagecaptioning,text2image}. Therefore, taking advantage of the cutting-edge deep generative models may contribute to addressing the wearable sensor data scarcity issue~\cite{DeepGenAccSynth}}.
\changezs{Another avenue of research is to utilize transfer learning, borrowing well-trained models from domains with high performing classifiers (i.e., images), and adapting them using a few samples of sensor data.}


\subsubsection{Data Quality and Missing Data}

\minorrevision{The quality of models is highly dependent on the quality of the training data. Many real-world collection scenarios introduce different sources of noise that degrade data quality, such as electromagnetic interference or uncertainty in task scheduling for devices that perform sampling~\cite{mDebugger}.
In addition to improving hardware systems, multiple algorithms have been proposed to clean or impute poor-quality data.} Data imputation is one of the most common methods to replace poor quality data or fill in missing data when sampling rates fluctuate greatly. For example, Cao \etal introduced a bi-directional recurrent neural network to impute time series data on the UCI localization dataset~\cite{NIPS_BRITS}. Luo \etal utilized a GAN to infer missing time series data~\cite{NIPS2018_GAN}. Saeed \etal proposed an adversarial autoencoder (AAE) framework to perform data imputation~\cite{autoencoder_graph}. \textbf{Opportunity:} To address this challenge, more research into automated methods for evaluating and quantifying the quality is needed to identify better, remove, and/or correct for poor quality data. \changezs{Additionally, it has been experimentally shown that deep neural networks have the ability to learn well even if trained with noisy data, given that the networks are large enough and the dataset is large enough~\cite{rolnick2017DL}. This motivates the need for HAR researchers to focus on other areas of importance, such as how to deploy larger models in real systems efficiently (Section~\ref{subsec:challenges_in_model_deployment}) and generate more data (Section~\ref{subsubsec:the_need_for_more_data}), which could potentially aid in solving this problem.}


\subsubsection{\changezs{Privacy Protection}}

%
\minorrevision{The privacy issue has become a concern among users~\cite{chen2021deep}. In general, the more inference potential a sensor has, the less willing a person is to agree to its data collection.
%
%
Multiple works have proposed privacy preservation methods while classifying human activities, including the replacement auto-encoder, the guardian, estimator, and neutralizer (GEN) architecture~\cite{AE6}, and the anonymizing autoencoder~\cite{AE7}.} For example, replacement auto-encoders learn to replace features of time-series data that correspond to sensitive inferences with values that correspond to non-sensitive inferences. Ultimately, these works obfuscate features that can identify the individual while preserving features common to each activity or movement. 
\changezs{
Federated learning is a trending approach to resolve privacy issues in learning problems~\cite{mothukuri2021survey, briggs2021review, sozinov2018human, Meta-HAR}. It can enable the collaborative learning of a global model without the need to expose users' raw data. 
~\citet{XIAO2021107338} realized a federated averaging method combined with a perceptive extraction network to improve the performance of the federated learning system.
~\citet{tu2021feddl} designed a dynamic
layer sharing scheme, which assisted the merging of local models to speed up the model convergence and achieved dynamic aggregation of models. 
~\citet{bettini2021personalized} presented a personalized semi-supervised federated learning method that built a global activity model and leveraged transfer learning for user personalization.
Besides,~\citet{gudur2021resource} implemented on-device federated learning using model distillation update and so-called weighted $\alpha$-updates strategies to resolve model heterogeneities on a resource-limited embedded system (Raspberry Pi), which proved its effectiveness and efficiency. 
}
\textbf{Opportunity:} 
\minorrevision{Blockchain is a new hot topic around the world. Blockchain, as a peer-to-peer network without the need for centralized authority, has been explored to facilitate the privacy-preserving data collection and sharing~\cite{8001731, bdiwi2018blockchain, 10.3389/fbloc.2020.497985, CHEN2021102011}.} 
%
The combination of federated learning and blockchain is also a potential solution towards privacy protection~\cite{nguyen2021federated} and is currently still in its very early stage. 
\minorrevision{More collaboration between ubiquitous computing community and networking community should be encouraged to prosper in-depth research in novel directions.}

\subsection{Challenges in Label Acquisition}

Labelled data is crucial for deep supervised learning. 
\minorrevision{
Image and audio data is generally easy to label by visual or aural confirmation. 
However, labelling human activities by looking at time series from HAR sensors is difficult or even impossible.
Therefore, label acquisition for HAR sensors generally requires additional sensing sources to provide video or audio data to determine the ground truth,
making label acquisition for HAR more labor-intensive. Moreover, accurate time synchronization between wearables and video/audio devices is challenging because different devices are equipped with independent (and often drifting) clocks. Several attempts have been made to address this issue, such as SyncWISE~\cite{SyncWISE} and~\cite{SynchronizationLex}. 
Two areas that require more research by the DL-HAR community are shortage in labelled data and difficulty in obtaining data from real-world scenarios.}

\subsubsection{Shortage of Labeled Data}
As annotating large quantities of data is expensive, there have been great efforts to develop various methods to reduce the need for annotation, including data augmentation, semi-supervised learning, weakly supervised learning, and active learning to overcome this challenge.
Semi-supervised learning utilizes both labelled data and unlabeled data to learn more generalizable feature representations. Zeng \etal presented two semi-supervised CNN methods that utilize unlabeled data during training: The convolutional encoder-decoder and the convolutional ladder network~\cite{semi1} and showed an 18\% higher F1-score using the convolutional ladder network on the ActiTracker dataset. 
Dmitrijs demonstrated on the SHL dataset, with a CNN and AAE architecture, that semi-supervised learning on unlabeled data could achieve high accuracy~\cite{AE422}. \changezs{Chen \etal proposed an encoder-decoder-based method that reduces distribution discrepancies between labelled and unlabeled data that arise due to differences in biology and behavior from different people while preserving the inherent similarities of different people performing the same task~\cite{chen2019people}.}
%


\minorrevision{Active learning is a special type of semi-supervised learning that selectively chooses unlabeled data based on an objective function that selects data with low prediction confidence for a human annotator to label. Recently, researchers have tried to combine DL approaches with active learning to benefit from establishing labels on the fly while leveraging the extraordinary classification capability of DL. Gudur \etal utilized active learning by combining a CNN with Bayesian techniques to represent model uncertainties (B-CNN)~\cite{DAL1}.
\changezs{Bettini \etal combined active learning and federated learning to proactively annotate the unlabeled sensor data and build personalized models in order to cope with data scarcity problem~\cite{bettini2021personalized}.}}
\textbf{Opportunity:} 
Though active learning has demonstrated that fewer labels are needed to build an effective deep neural network model, 
a real-world study with time-cost analysis would better demonstrate the benefits of active learning. Moreover, given the many existing labelled datasets, another area of opportunity is developing methods that \minorrevision{leverage characteristics of labelled datasets to generate labels for unlabeled datasets such as transfer learning or pseudo-label method~\cite{pseudolabel}.}








\subsubsection{Issues of In-the-field Dataset}
\minorrevision{
Traditionally, HAR research has been conducted primarily in lab. Recently, HAR research has been moving towards in-field experiments. Unlike in-lab settings, where the ground truth can be captured by surveillance cameras, in-field experiments may have subjects moving around in daily life, where static camera deployment is not sufficient any more. Alharbi \etal used wearable cameras placed at the wrist, chest, and shoulder to record subject's activities as they moved around outside of a lab setting~\cite{barriers} and studied the feasibility of wearable cameras.} 
\textbf{Opportunity:} More research in leveraging human-in-the-loop to provide in-field labelling is required to generate more robust datasets for in situ activities. 
\minorrevision{Besides, one possible solution is to utilize existing in-the-field human activity video datasets and cross-modal deep generative models. If high-fidelity synthetic wearable sensor data can be generated from the available real-world video datasets (such as Stanford-ECM dataset~\cite{stanford_ecm}) or online video corpus, it may help alleviate the in-the-field data scarcity issue.} 
Additionally, there are opportunities for semi-supervised learning methods that leverage the sparse labels provided by humans-in-the-loop to generate high-quality labels for the rest of the dataset.





\subsection{Challenges in Modeling}

\changezs{
In this section, we discuss the challenges and opportunities in the modelling process in several aspects, including data segmentation, semantically complex activity recognition, model generalizability, as well as model robustness. 
}

\subsubsection{Data Segmentation}

As discussed in~\cite{review_Plotz}, many methods segment time series using traditional static sliding window methods. A static time window may either be too large, \minorrevision{capturing more than necessary to detect certain activities, or too small and not capturing enough series to detect long movements}. Recently, researchers have been looking to segment time series data more optimally. Zhang \etal used reinforcement learning to find more optimal activity segments to boost HAR performance~\cite{RL1}. 
\change{\citet{qian2021weakly} proposed weakly-supervised sensor-based activity segmentation and recognition method.}
\textbf{Opportunity:} More experimentation and research into dynamic activity segments or methods that leverage both short term and long term features (i.e., wavelets) are needed to create robust models at all timescales. \changezs{While neural networks such as RNNs and LSTMs can model time series data with flexible time scales and automatically learn relevant features, their inherent issues such as exploding/vanishing gradients and training difficulty, make widespread adoption difficult. As such, more research into other methods that account for these issues is necessary.}

\subsubsection{Semantically Complex Activity Recognition}

\minorrevision{Current HAR methods achieve high performance for simple activities such as running. However, complex activities such as eating, which can involve a variety of movements, remain difficult.} To tackle this challenge, Kyritsis \etal break down complex gestures into a series of simpler (atomic) gestures that, when combined, form the complex gesture~\cite{eatingmicromovement2019}. Liu \etal propose a hierarchical architecture that constructs high-level human activities from low-level activities~\cite{hierarchical2016}. Peng \etal proposes AROMA, a complex human activity recognition method that leverages deep multi-task learning to learn simple activities that make up more complex \minorrevision{movements}~\cite{aroma}. \textbf{Opportunity:} Though hierarchical methods have been introduced for various complex tasks, there are still opportunities for improvements. Additionally, novel black-box approaches to complex task recognition, where individual steps in complex actions are automatically learned and accounted for rather than specifically identified or labelled by designers, have yet to be fully explored. \minorrevision{Such a paradigm is perfectly suitable for deep learning because neural networks function on a similar principle.} Besides, graph neural network can also be explored to model the hierarchical structure of simple-to-complex human activities~\cite{gnn_survey}. 



\subsubsection{Model Generalizability}

A model has high generalizability when it performs well on data that it has never seen before. Overfitting occurs when it performs well on training data but poorly on new data. \minorrevision{Recently, many efforts have been put into improving the generalizability of models in HAR~\cite{Stisen, Heterogeneities2, Heterogeneities3}.} 
\minorrevision{
Most research on generalizability in HAR has been focused on creating models that can generalize to a larger population, which often requires a large amount of data and high model complexity. In scenarios where high model complexity and data are not bottlenecks, DL-based HAR generally outperforms and generalize better than other types of methods. In scenarios where data or model complexity is limited, DL-based methods must utilize available data more efficiently or adapt to the specific scenario online.} 
For instance, Siirtola and Röning propose an online incremental learning approach that continuously adapts the model with the user's individual data as it comes in~\cite{incremental}. \change{\citet{qian2021latent} introduce Generalizable Independent Latent Excitation (GILE), which greatly enhances the cross-person generalization capability of the model.}
\changezs{\textbf{Opportunity:} \minorrevision{An avenue of generalizability that has yet to be fully explored are new training methods that can adapt and learn predictors across multiple environments, such as invariant risk minimization~\cite{arjovsky2020invariant} or federated learning methods~\cite{konevcny2015federated}.} Incorporating these areas into DL-based HAR could not only improve the generalizability of HAR models but accomplish this in a model-agnostic way.}


\subsubsection{Model Robustness}

\changezs{
A key issue that the community is paying increasing attention to is model robustness and reliability~\cite{multisensor_review, ahad2021sensor}. 
%
%
%
%
One common way to improve robustness is to leverage the benefits of multiple types of sensors together to create {multi-sensory systems}~\cite{abedin2021attend,Physical_Activity2021,sd_multisensory,ijcai2018-859,RL1,sena2021human}.
%
\citet{Physical_Activity2021} has proposed an architecture called DeepFusionHAR to incorporate the handcrafted features and deep learning extracted features from multiple sensors to detect daily life and sports activities.
\citet{sd_multisensory} proposed a multi-sensory approach for basic and complex human activity recognition that uses built-in sensors from smartphones and smartwatches to classify 20 complex actions and five basic actions.
\citet{ijcai2018-859} demonstrated a mobile \minorrevision{application on a multi-sensor} mobile platform for daily living activity classification using a combination of accelerometer, gyroscope, magnetometer, microphone, and GPS. 
Multi-sensory networks in some cases are integrated with attention modules to learn the most representative and discriminative sensor modality to distinguish human activities~\cite{RL1}.} 
\changezs{\textbf{Opportunity:} While there are works that utilize multiple sensors to improve robustness, \minorrevision{they} require users to \minorrevision{wear} or have access to all of the sensors they utilize. An exciting new direction is to create generalized frameworks that can adaptively utilize data from whatever sensors happen to be available, such as a smart home intelligence system~\cite{xiadia2021}. \minorrevision{For this direction, deep learning methods seem more suitable than classical machine learning methods because neural networks can be more easily tuned and adapted to different domains (i.e., different sensors) than rigid classical models, just by tuning weights or by mixing and matching different layers or embeddings.} Creating such systems would not only greatly improve the practicality of HAR-based systems but would also contribute significantly to general artificial intelligence.}







\subsection{Challenges in Model Deployment}

\label{subsec:challenges_in_model_deployment}

\change{\minorrevision{There are several works focusing on deploying deep-learning-based HAR on mobile platforms.}~\citet{deepx} proposes a SOC-based architecture for optimizing deep neural network inference, while~\citet{DeepEar} and \citet{MobiRNN} utilize the smartphone's digital signal processor (DSP) and mobile GPU to improve inference time and reduce power consumption.~\citet{DeepSense} propose a lightweight CNN and RNN-based system that accounts for noisy sensor readings from smartphones and automatically learns local and global features between sensor windows to improve performance.}

\change{The second class of works focus on reducing the complexity of neural networks so that they can run on resource-limited mobile platforms.~\citet{Sparsification} reduces the amount of computation required at each layer by encoding layers into a lower-dimensional space.~\citet{Binarized-BLSTM-RNN} reduces computation by utilizing binary weights rather than fixed-point or floating weights.}

\change{Emerging trends in deploying neural networks include offloading computation onto application-specific integrated circuits (ASIC) or lower power consumption microcontrollers.~\citet{bhat2019asic, wang2008asic} developed custom integrated circuits and hardware accelerators that perform the entire HAR pipeline with significantly lower power consumption than mobile or GPU-based platforms. The downside to ASICs is that they cannot be reconfigured for other types of tasks.~\citet{islam2020zygarde} present an architecture for embedded systems that dynamically schedules DNN inference tasks to improve inference time and accuracy.}

\change{\textbf{Opportunity:} Though there are works that explore the deployment of DNNs practical systems, more research is needed for society to fully benefit from the advances in DNNs for HAR. Many of the works discussed leverage a single platform (i.e., either a smartphone or ASIC), but there are still many opportunities for improving the practical use of HAR by exploring intelligent ways to partition computation across the cloud, mobile platforms, and other edge devices. DNN-based HAR systems can largely benefit by incorporating methodologies proposed by works such as~\cite{xia2021csafe, xia2019improving, godoy2018paws, nie2020spiders, nie2021spidersplus, nie2020facial, chandrasekaran2016seus, xia2018smartphone}, that carefully partition computation and data across multiple devices and the cloud.}

\change{
Lane \etal performed a small-scale exploration into the performance of DNNs for HAR applications on mobile platforms in various configurations, including utilizing the phone's CPU and DSP and offloading computation onto remote devices~\cite{lane2015mobile}.} 
\change{This work demonstrates that mobile devices running DNN inference can scale gracefully across different compute resources available to the mobile platform and also supports the need for more research into optimal strategies for partitioning DNN inference across mobile and edge systems to improve latency, reduce power consumption, and increase the complexity of the DNNs serviceable to wearable platforms.
}







\section{Conclusions} \label{sec:conclusions}



Human activity recognition in wearables has provided us with many conveniences and avenues to monitor and improve our life quality. AI and ML have played a vital role in enabling HAR in wearables. 
In recent years, DL has pushed the boundary of wearables-based HAR, bringing activity recognition performance to an all-time high. 
\change{In this paper, we provided our answers to the three research questions we proposed in Section 
\ref{sec:method}. We firstly gave an overall picture of the real-life applications, mainstream sensors, and popular public datasets of HAR. Then we gave a review of the advances of the deep learning approaches used in the field of wearable HAR and provided guidelines and insights about how to choose an appropriate DL approach after comparing the advantages and disadvantages of them. At last, we discussed the current road blockers in three aspects---data-wise, label-wise, and model-wise---for each of which we provide potential opportunities.
We further identify the open challenges and finally provide suggestions for future avenues of research in this field. 
By categorizing and summarizing existing works that apply DL approaches to wearable sensor-based HAR, we aim to provide new engineers and researchers entering this field an overall picture of the existing research work and remaining challenges. We would also like to benefit experienced researchers by analyzing and discussing the developing trends, major barriers, cutting-edge frontiers, and potential future directions.
}


\section*{Acknowledgments}
Special thanks to Haik Kalamtarian and Krystina Neuman for their valuable feedback.




\end{document}